\documentclass[final]{IEEEtran} 
\usepackage{amsthm,amssymb,graphicx,multirow,amsmath,color,amsfonts}
\usepackage[update,prepend]{epstopdf}
\usepackage[noadjust]{cite}
\usepackage[latin1]{inputenc}
\usepackage{tikz}
\usetikzlibrary{angles, quotes}
\usepackage{bbm} 
\usepackage{pdfpages}
\usepackage{tabulary}
\usepackage{multirow}
\usepackage{comment}
\usepackage{dsfont}
\usetikzlibrary{calc}
\usepackage{ amssymb }
\usepackage[mathscr]{eucal}
\usepackage[caption=false]{subfig}
\usepackage{float}

\newtheorem{corollary}{Corollary}

\def\nb0{{\mathbf{0}}}
\def\nb1{{\mathbf{1}}}

\newtheorem{theorem}{Theorem}

\def\argmax{\operatorname{arg~max}}
\def\E{\mathbb{E}}

\def\P{\mathbb{P}}

\def\R{\mathbb{R}}

\allowdisplaybreaks 

\begin{document}
\graphicspath{{./Figures/}}
\title{Coverage Analysis for UAV-Assisted Cellular Networks in Rural Areas}
\author{
Maurilio Matracia, \em Student Member, IEEE, \normalfont Mustafa A. Kishk, \em Member, IEEE, \\ \normalfont and Mohamed-Slim Alouini, \em Fellow, IEEE 
\thanks{The authors are with Computer, Electrical, and Mathematical Sciences
and Engineering (CEMSE) Division at King Abdullah University of Science
and Technology (KAUST), Thuwal, 23955-6900, KSA (email: $\{$maurilio.matracia; mustafa.kishk; slim.alouini$\}$@kaust.edu.sa). 
}}

\maketitle

\begin{abstract}
Despite coverage enhancement in rural areas is one of the main requirements in next generations of wireless networks (i.e., 5G and 6G), the low expected profit prevents telecommunication providers from investing in such sparsely populated areas.
Hence, it is required to design and deploy cost efficient alternatives for extending the cellular infrastructure to these regions.
A concrete mathematical model that characterizes and clearly captures the aforementioned problem might be a key-enabler for studying the efficiency of any potential solution. 
Unfortunately, the commonly used mathematical tools that model large scale wireless networks are not designed to capture the unfairness, in terms of cellular coverage, suffered by exurban and rural areas. 
In big cities, in fact, cellular deployment is essentially capacity driven and thus cellular base station densities are maximum in the town centers and decline when getting far from them.
In this paper, a new stochastic geometry-based model is implemented in order to show the coverage spatial variation among urban, suburban, and exurban settlements. 
Indeed, by implementing inhomogeneous Poisson point processes (PPPs) it is possible to study the performance metrics in a realistic scenario where terrestrial base stations (TBSs) are clustered around the urban center while  outer aerial base stations (ABSs) are uniformly distributed outside an urban exclusion zone. 
Based on this, our simulation results can quantify the improvement, in terms of coverage probability, that even a surprisingly low density of ABSs can bring to peripheral regions depending on the extension of the exclusion zone, enabling us to draw insightful considerations.
\end{abstract}


\section{Introduction} \label{sec:Intro}
Even though wireless connectivity is rapidly spreading all around the world, the number of uncovered users living in rural and low income areas is still large because telecommunication operators are not motivated to invest in low return on investment (ROI) rate zones \cite{Chiaraviglio19}. 
Therefore, appropriate countermeasures are needed in order to even the connectivity opportunities of all users, no matter the type of area they live in. \par
During the last decade, the number of use cases and applications of unmanned aerial vehicles (UAVs), especially drones, dramatically increased \cite{Mozaffari19, fotouhi2019survey} despite their strong limitations in terms of payload capability and flight time. 
Particular research interest now regards the introduction of UAV-mounted base stations (BSs) in order to serve ground users \cite{Wu18, Zeng16}. 
This is motivated by the fact that increasing the altitude of the BSs eventually leads to higher probability of establishing a line-of-sight (LoS) channel, which is more efficient than its non-LoS (NLoS) counterpart. 
Finally, low cost, high mobility, and fast deployability of drones make them appropriate for covering large regions with low traffic demand density such as rural areas, apart from the fact that they could be used for other applications such as precision farming \cite{Vasudevan16} or precision agriculture \cite{Muchiri16} and even delivery of small and light packages \cite{Dorling2017}. \par
One of the most popular mathematical tools that are widely used to analyze cellular networks is stochastic geometry.
However, there is always an underlying assumption of homogeneity, which does not capture the spatial variation of coverage between different environments such as urban or rural.
In this paper, we develop a stochastic geometry based framework that captures such variation while deploying ABSs to enhance the coverage only where required, which is mainly rural and remote areas.
More details on the contributions of this paper will be provided in Sec. \ref{sec:contributions}. 
But first, we enlist some of the most relevant literature in the next subsection.

\subsection{Related Work}
This subsection provides a concise summary of the related papers belonging to two general fields of study: (i) coverage enhancement in rural areas and (ii) stochastic geometry-based
analysis of UAV networks. 

\subsubsection{Coverage Enhancement in Rural Areas} Several papers suggested various solutions for ensuring better services to rural users.
In \cite{Yaacoub19}, for instance, an overview of the main smart and cost-aware fronthaul and backhaul solutions for rural connectivity has been provided for the most common scenarios.
Similarly, authors in \cite{khalil2019comparative} compared relevant communication paradigms to evaluate their effectiveness in rural environments, based on their network architectures, performance parameters, and deployability.
On the other hand, authors in \cite{Chiaraviglio17} firstly developed a simplified model to evaluate the required subscription fees for users living in rural and low income environments in order to amortize the costs of a 5G architecture, eventually proving that the UAV-based solution would be effective. 
Next, an innovative algorithm was proposed in \cite{Chiaraviglio19} in order to optimize the selection of the ground sites needed for the UAVs and their optical fiber links, taking into account both coverage and energy constraints. 
However, since there are still large low-income rural regions around the globe that need very basic connectivity, works such as \cite{Anusha17} mostly discussed previous wireless technologies (including 2G and 3G), primarily with respect to their performance in cost. 
In \cite{khalil2017feasibility}, TV band white space (TVWS) with 5G infrastructure is suggested as a supplement for rural communications, since it may be cost-effective from a service provider's perspective. \par
Another important field of study regards the improvement of coverage and capacity for rural vehicular users, which has been evaluated in case of deployment of high-altitude platforms (HAPs) in \cite{popoola2020capacity}. 
On the other side, studies such as \cite{Hasan13} proposed software defined networks (SDNs) in order to compensate the unpredictability that characterizes rural zones by efficiently managing their wireless networks.
Finally, it is believed  that dense deployment of low Earth orbit (LEO) satellites can lead to improving the coverage in various rural and remote regions \cite{talgat2020nearest, talgat2020}.

\subsubsection{Stochastic Geometry for UAV Networks} Nowadays, stochastic geometry is recognized as an effective mathematical tool to model and analyze various performance metrics of wireless networks \cite{ElSawy13, ElSawi17,Hou_NOMA,Alzenad18,Armeniakos20,Alzenad19,arshad2018integrating,Qin2020,kishk20203}.
The performances of UAV networks and UAV-assisted terrestrial cellular networks (typically referred to as vertical heterogeneous networks or VHetNets) have been evaluated by means of stochastic geometry in works such as \cite{Alzenad18,Armeniakos20} and \cite{Alzenad19}, respectively.
In particular, \cite{Alzenad19} assumes a setup where both TBSs and ABS are uniformly distributed over their respective horizontal planes.
Another interesting setup is proposed by authors in \cite{arshad2018integrating}, where the network architecture consists of three tiers (macro and small TBSs supported by ABSs).
As shown in \cite{Qin2020}, stochastic geometry approaches are effective also for capturing how the coverage probability depends on the spatial distribution of the charging stations.
While all the aforementioned works referred to untethered UAVs (U-UAVs), in \cite{kishk20203} an innovative stochastic geometry approach is proposed to derive the probability distribution of the minimum inclination angle of the wires supplying power and data to tethered UAVs (T-UAVs).

\subsection{Contributions} \label{sec:contributions}
The contributions of this paper can be summarized as follows: \\
$\bullet$ This paper provides the first stochastic geometry-based framework  specifically designed to analyze the downlink performance of large scale wireless networks while capturing the performance variation among different types of environments: urban, suburban, exurban, and rural areas; \\
$\bullet$ The unfairness (in terms of coverage probability) experienced by users located
far from town centers is captured by modeling the locations of terrestrial base stations using inhomogeneous PPP; \\
$\bullet$ Given the inherent challenges in analyzing inhomogeneous PPP-modeled networks, we derive various novel distance distributions in order to compute the performance of mobile users as a function of their distance from the town/city center; \\ 
$\bullet$ Unlike existing works, where ABSs are typically assumed to be deployed everywhere, we propose a realistic setup where ABSs are sent only in regions that suffer from lower performance. 
This is achieved by introducing an exclusion zone, in which the deployment of ABSs is not needed since existing TBSs are already providing sufficient coverage probability. 
We compute the coverage probability for this novel setup and show how the degradation in the performance, as the users move away from the town or the city center, can be lifted in a cost efficient manner by deploying ABSs only where needed.

\section{System Model} \label{sec:S}
\subsection{Network Modeling}
Letting the types of BSs $A$, $L$, $N$, $T$, $M$, $B$, $C$, and $Q$ be defined as in table \ref{tab:subscripts}, their use as subscripts or superscripts will characterize each quantity accordingly.

\begin{table*}[h!] 
  \begin{center}
    \caption{BSs' Subscripts and Superscripts}
    \label{tab:subscripts}
     \begin{tabular}{|c|c|c|} \hline
      \textbf{Subscript or Superscript} & \textbf{Description} & \textbf{Definition}  \\
      \hline
      ${A}$ & ABSs & $-$ \\ \hline 
      ${L}$ & LoS ABSs & $-$ \\ \hline
      ${N}$ & NLoS ABSs & $-$ \\ \hline 
      ${T}$ & TBSs & $-$ \\ \hline
      ${M}$ & LoS exclusive or NLoS ABSs & $M\in\{L,N\}$  \\ \hline
      ${Q}$ & Generic type of BS & $Q\in\{L,N,T\}$ \\ \hline
      ${B}$ & Type of tagged BS & $B\in\{L,N,T\},\, K_{B,W^*}> K_{Q,W_i},\, \forall Q, \, \forall W_i\neq W^*$ \\ \hline
      ${C}$ & Type of interfering BSs & $C\in\{L,N,T\},\, K_{C,W_i}<K_{B,W^*},\,\forall W_i\neq W^*$ \\ \hline
      
    \end{tabular}  
  \end{center} 
\end{table*} 

\begin{figure*}[t]
\centering
\includegraphics[width=1.8\columnwidth, trim={0cm 0.3cm 0cm 0cm},clip]{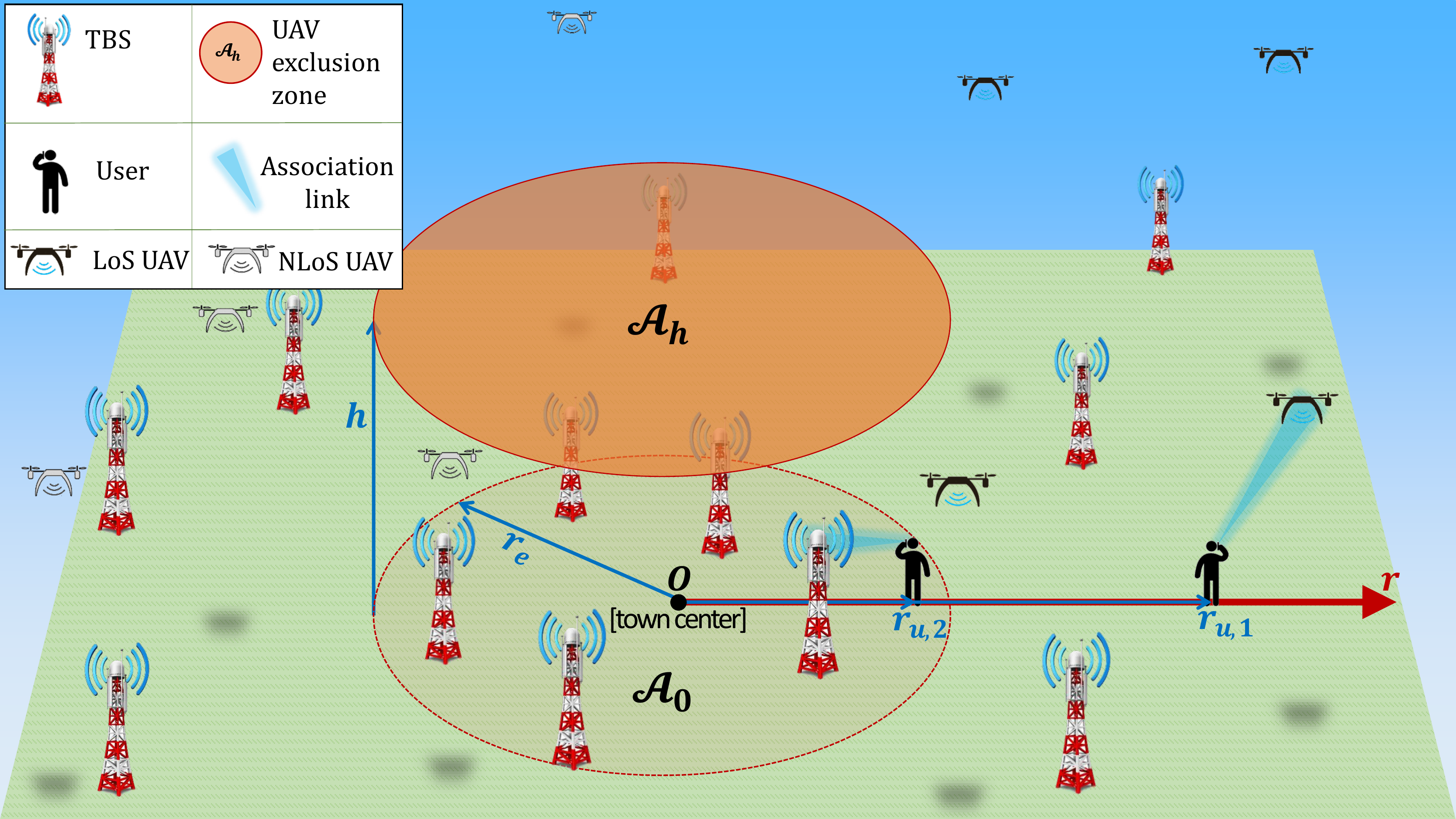}
\caption{Schematic representation of the setup considered: each user $U_i$ is located at a distance $r_{u,i}$ (with $i\in\mathbb{N}^+$) from the origin and associates to the BS that provides the maximum average received power. 
The exclusion zone is showed in red at altitude $h$, which is common to all the UAVs.}
\label{fig:system}
\end{figure*}

By introducing a polar coordinate system $(\theta,r)$ centered around the town center, we introduce a downlink cellular network in which all the TBSs are located according to an inhomogeneous PPP $\Phi_T\equiv\{Y_i\}\subset\R^2$ of density $\tilde{\lambda}_T(y)=\mathcal{G}_T(\|y\|)\,\lambda_T$, where $\mathcal{G}_T(\|y\|)$ represents the probability density function (PDF) describing the planar distribution of the TBSs. 
To describe our scenario, $\mathcal{G}_T(\|y\|)$ is assumed to be a monotonically decreasing function. 
In addition, we consider the deployment of ABSs at altitude $h$ (assumed to be constant, for simplicity) according to a PPP $\Phi_A$ with density $\lambda_A$ outside an exclusion zone of radius $r_e$ centered at around the origin (see Fig. \ref{fig:system}). 
Note that the PPP assumption is widely accepted for modeling cellular networks and has been empirically verified in many studies (e.g., \cite{Andrews11,Guo13,Lu15}). \par
We will derive mathematical expressions for the coverage probability considering that the signals transmitted by any $Q$ BS have fixed constant transmit power $p_Q$, and experience standard power-law path loss propagation model with path loss exponent $\alpha_Q>2$.
For simplicity, it is also assumed that random channel effects are incorporated by multiplicative random values that correspond to Nakagami fading gains annotated by $H$ for the desired signal and $G_{Q,W_i}$ for each interferer of type $Q$ located at the point $W_i$ of the respective point process.
For the terrestrial network model, we consider a scenario where $\tilde{\lambda}_T$ is only function of the horizontal distance $r$ from the origin, which means that the system is radially-symmetric and hence isotropic (i.e., rotation-invariant when rotated around the origin \cite{haenggi2012stochastic}). As a result, if the typical user is located at a generic point $U=(\theta_u, \,r_u)$, we can arbitrarily put $\theta_u=0$ for simplicity\footnote{Indeed, even assuming $\theta_u\neq0$ it would still be possible to perform a rotation of the axis around the origin such that the new angle of the user is zero.}. \par
Regarding the aerial network model, UAVs are deployed starting from a distance $r_e$ from the origin, according to a homogeneous PPP with density $\lambda_A$ on a horizontal plane at fixed altitude $h$.
Note that this assumption is very realistic, since it has been shown in \cite{Chetlur17bpp} that the coverage probability depends on the average value of ABSs' altitude rather than on its distribution.
Hereby, backhaul links of the ABSs are considered ideal, meaning that they have sufficient bandwidth and no errors.
Practical solutions for backhaul in rural areas might rely on LEO satellites, TBSs, or even tethered UAVs, as suggested in \cite{Kishk20magazine}.
However, this paper mainly  focuses on the use of ABSs for providing more a reliable access link to the typical user.

\subsection{Channel Modeling}
This subsection aims to characterize both the terrestrial and the aerial wireless channels.
Let $m$ and $\eta$ respectively denote the Nakagami-m shape parameter and the mean additional transmission losses.
Introducing $\xi_Q=\eta_Q\,p_Q$, then the terrestrial and the air-to-ground (A2G) channels can be characterized as follows.

\subsubsection{Terrestrial Channel}
For the terrestrial links, we consider a standard power-law path-loss model given by $\Lambda_T(z)=\eta_T\,z^{\alpha_T}$ for any TBS located at generic distance $z$ from the user, where $L$ denotes the path loss.
Furthermore, we assume that the terrestrial links experience small-scale fading in the form of Nakagami-m distribution with shape parameter $m_T$, and the channel fading power gains $G_{T,Y_i}$ follow a Gamma distribution with PDF given by
\begin{align}
f_{G_{T,Y_i}}(g)=\frac{m_T^{m_T}\,g^{m_T-1}}{\Gamma(m_T)}\,e^{-m_T\,g},\,\,\,\,\forall Y_i\in \Phi_T,    
\end{align} 
where $Y_i$ indicates the location of the $i$-th TBS and $\Gamma(m)=\int\limits_0^\infty x^{m-1}\,e^{-x}\,{\rm d}x$ is the Gamma function.
Hence, the received power at the typical user from a TBSs located at $Y_i$ is given by $K_T=\xi_T\,G_{T,Y_i}\,Z_T^{-\alpha_T}$, where $Z_T$ represent the random horizontal distance between the typical user and the TBS considered.

\subsubsection{A2G Channel} 
Small-scale fadings on NLoS and LoS links are usually Rayleigh or Rician distributed, respectively.
However, the Nakagami-m distribution with shape parameter\footnote{Recall that the respective scale parameter is equal to the reciprocal of the shape parameter.}   $m=\frac{(\mathcal{K}+1)^2}{2\mathcal{K}+1}$ approximates the Rician distribution with factor $\mathcal{K}$ \cite{Alzenad19}.
Furthermore, independent small-scale fading will be considered on each link for NLoS and LoS transmissions, which occur with specific probabilities depending on the height and density of the buildings, the type of environment, and the elevation angle.
Indeed, according to \cite{al2014optimal} we introduce the LoS probability as
\begin{align}
\mathcal{P}_L(z)=\frac{1}{1+\mathcal{S}_a\,\exp\Big(-\mathcal{S}_b\,\left(\frac{180}{\pi}\,\tan^{-1}\left(\frac{h}{z}\right)-\mathcal{S}_a\right)\Big)},
\end{align}
where $\mathcal{S}_a$ and $\mathcal{S}_b$ are the so-called s-curve parameters and are environmental constants, $z$ is the horizontal component of the Euclidean distance between the typical user and the ABSs.
For the sake of an easier tractability, we assume $\mathcal{S}_a$ and $\mathcal{S}_b$ to be uniform over the entire environment considered.
Trivially, the NLoS probability is the complement to unity of the LoS one, that is $\mathcal{P}_N(z)=1-\mathcal{P}_L(z)$.
Each ABS is either in LoS or NLoS condition with the user, independently of the other ABSs.
From the user's perspective, therefore, it follows that $\Phi_A=\Phi_L\cup\Phi_N$, where each point of $\Phi_A$ is mapped into $\Phi_L$ if the respective ABS is in LoS condition with the user and into $\Phi_N$ otherwise. \par
The channel fading power gains for either LoS or NLoS ABSs, $G_{M,X_i}$, follow similar distributions to their terrestrial counterpart, i.e.
\begin{align}
f_{G_{M,X_i}}(g)=\frac{m_M^{m_M}\,g^{m_M-1}}{\Gamma(m_M)}\,e^{-m_M\,g},\,\,\,\,\forall X_i\in \Phi_M,
\end{align}
where $M\in\{L,N\}$ and $X_i$ indicates the location of the $i$-th ABS and $m$ denotes the Nakagami-m shape parameter. 
The received power at the typical user from an ABS located at point $X_i$ can be computed as
\begin{align}
K_{M,X_i}=\xi_M\,G_{M,X_i}\,D_{M,X_i}^{-\alpha_M}, \,\,\,\,\forall X_i\in \Phi_M,
\end{align}
where $D_{M,X_i}$'s are the random variables (RVs) identifying the Euclidean distances between any ABS located at the point $X_i$ and the typical user. 

\subsection{Association Policy}
In this paper, the strongest average received power association rule is adopted as in \cite{Alzenad19}, that is, the user connects to the BS providing the highest average received power.
This, however, does not exclude the possibility of having an interfering BS that provides the highest instantaneous received power.
Also, note that due to eventual differences in terms of path-loss exponent, mean additional transmit losses, and transmit power, the serving BS may be farther than some interfering BSs from other tiers. 
However, it is evident that the serving BS is always the closest within its tier. 
Finally, no bias cell association is adopted and hence the expected values of the fading gains over all the sets of BSs (i.e., $\mathbb{E}[G_{Q,W_i}], \forall W_i\in\Phi_Q$) are assumed unitary.
Because of this, the location of the tagged BS will be provided by the maximum product $\xi_Q\,D_{Q,W_i}^{\alpha_Q}$, that is 
\begin{align}
W^*=\underset{\underset{Q\in\{L,N,T\}}{W_i\in\Phi_Q}}{\argmax} \,(\xi_Q\,D_{Q,W_i}^{-\alpha_Q})\,.
\end{align}

\subsection{Interference and Signal to Interference plus Noise Ratio ($\rm SINR$)} 
Being $W^*\in\Phi_B$ the location of the serving BS, the instantaneous SINR at the typical user can be expressed as 
\begin{align}
    {\rm SINR}=\frac{K_{B,W^*}}{\sigma_n^2+I}, 
\end{align}
where $\sigma_n^2$ is the additive white Gaussian noise (AWGN) power and $I$ is the aggregate interference power.
Since $C$ denotes the type of each interfering BS, then the RV $I$ can be introduced as
\begin{align}
    I=\sum\limits_{C\in\{L,N,T\}}\,\sum\limits_{\substack{W_i\in\Phi_C \\ {W_i\neq W^*}}} K_{C,W_i}.
\end{align}

\subsection{Coverage Probability}
The coverage probability is defined as the complementary cumulative distribution function (CCDF) of the SINR evaluated at a designated threshold $\tau$, that is
\begin{align}
    P_c=\P(\text{SINR}>\tau).
\end{align}

\section{Performance Analysis} \label{sec:PA}
In this section, the distributions of the distance to the closest BS, the SINR, the association probabilities, and the conditional Laplace transforms of the interference will be derived for each type of tagged BS in order to obtain the approximate coverage probability.

\subsection{Distance to the Nearest TBS}
Intuitively, the coverage probability is function of the distance of the tagged BS.
In order to derive the final expression of the coverage probability, for each type of BS we characterize the distribution of the horizontal distance between the user and the closest transmitter by computing the cumulative distribution function (CDF) and consequently deriving the PDF.
\begin{theorem}  \label{thm:CDF_T}
Let  $(\beta,z)$ be the polar coordinate system centered at the user, then the CDF of the random horizontal distance\footnote{Whenever not specified, we always refer to the distance from the user, around whom the polar coordinate system is centered.
Note also that for any TBS the Euclidean distance equals its horizontal projection.} $Z_T$ between the user and the closest TBS in an inhomogeneous PPP with density $\tilde{\lambda}_T(r)=\mathcal{G}_T(r)\,\lambda_T$ is given by
\begin{align}
  F_{Z_T}(z)=&
  1-\exp\bigg(-\lambda_T\int\limits_0^z\int\limits_{-\pi}^\pi \mathcal{G}_T(r(\beta,z'))\,r(\beta,z')\,{\rm d}\beta\,{\rm d}z'\bigg), \label{eq:CDF_T}
\end{align} 
where $r(\beta, z')=\sqrt{r_u^2+z'^2-2\,z'\,r_u\,\cos\beta}$ is the horizontal distance to the town center expressed in the polar coordinate system centered around the typical user, and $r_u$ is the distance between the typical user and the town center.
\end{theorem}
\begin{IEEEproof}See Appendix \ref{appx:CDF_T}. \end{IEEEproof}

\begin{corollary}\label{cor:PDF_T}
The PDF of the distance between the user and the serving TBS can be computed as
\begin{align}
f_{Z_T}(z)=& 
\bar{F}_{Z_T}(z)\,\lambda_T\,\int\limits_{-\pi}^\pi \mathcal{G}_T(r(\beta,z))\,r(\beta,z)\,{\rm d}\beta,
\end{align}
where $\bar{F}_{Z_T}(z)$ denotes the CCDF of the RV $Z_T$. 
\end{corollary}
\begin{IEEEproof} The result follows directly by taking the derivative of the expression in (\ref{eq:CDF_T}). \end{IEEEproof}

\begin{figure}[t]
\centering
\subfloat{\includegraphics[width=1.29\columnwidth, trim={3cm 0cm 1cm 0cm},clip]{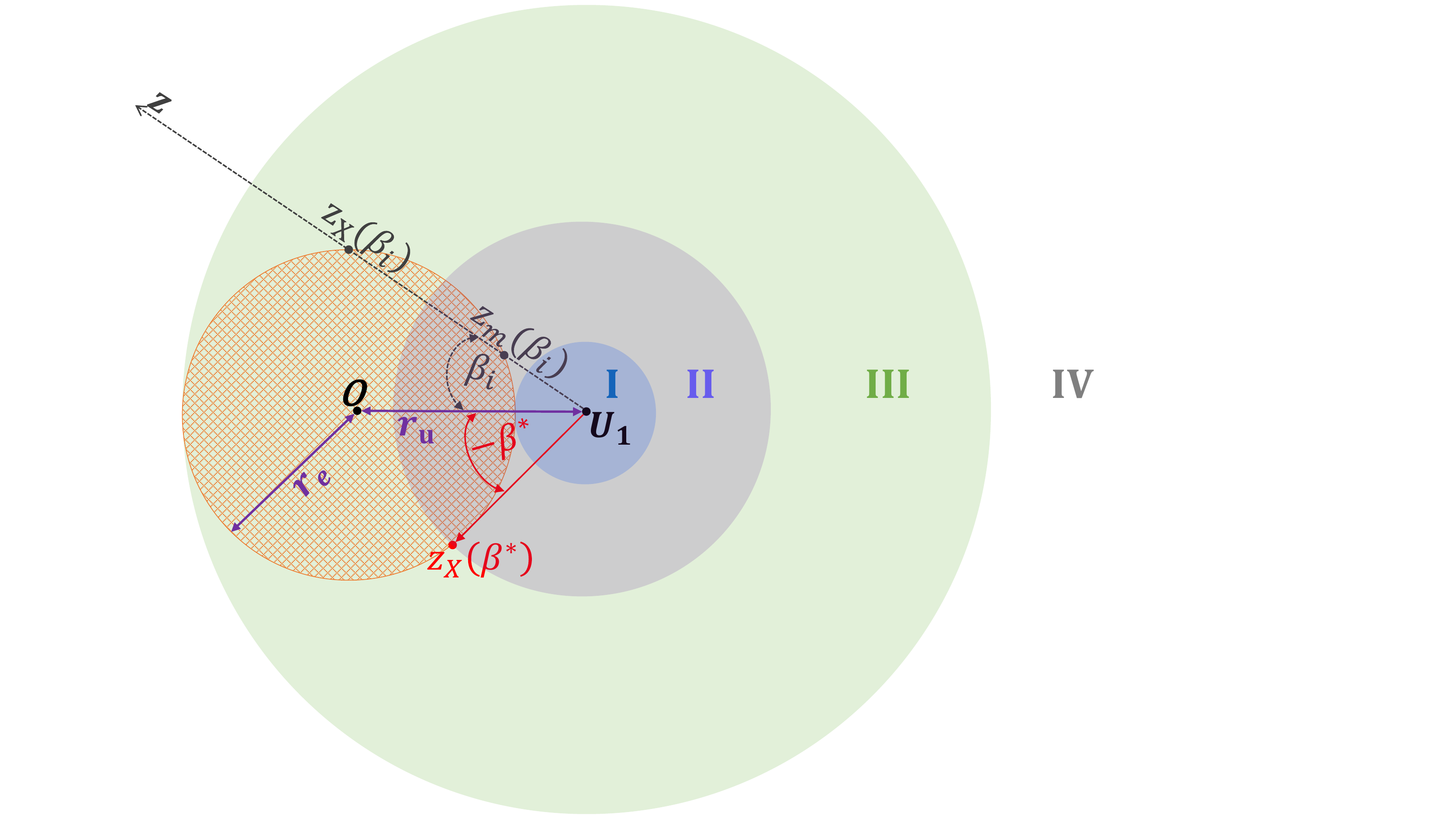}  \label{fig:regions1}}\\
\subfloat{\includegraphics[width=1.19\columnwidth, trim={5cm 0cm 0.5cm 0cm},clip]{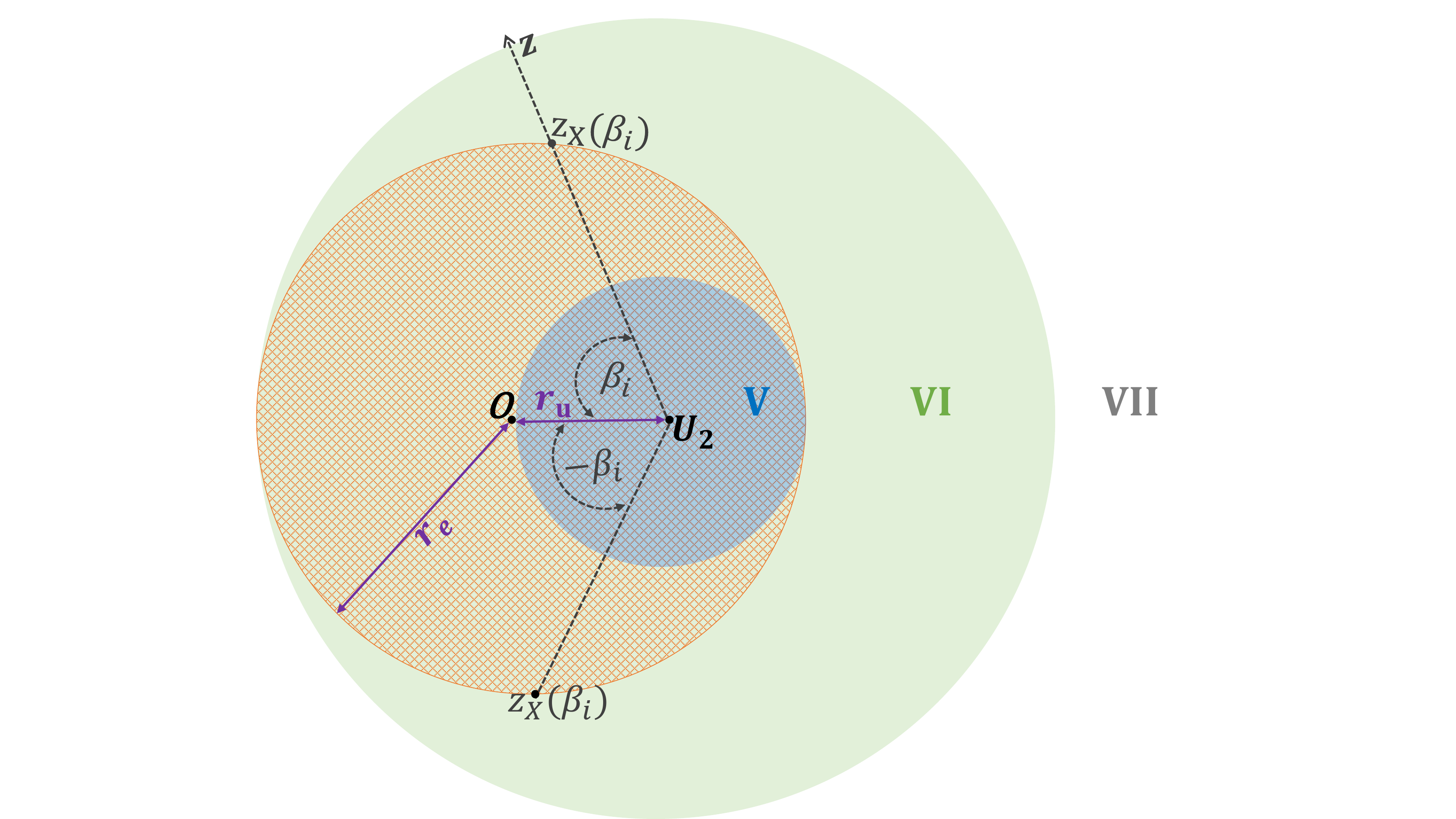}  \label{fig:regions2}}
\caption{Schematic representation of the seven different regions formed when introducing the vacancy (i.e., the grid shown in gold) for ABSs.
The four regions on the top correspond to the case with $r_u>r_e$, whereas the remaining three regions refer to the case $r_u\leq r_e$.}
\label{fig:regions}
\end{figure}

\begin{theorem} \label{thm:CDF_M}
When deploying the aerial network in the proposed scenario, the following seven regions (four if the user is outside the exclusion zone and three otherwise) need to be identified according to  Fig. \ref{fig:regions} in order to describe the distribution of the minimum horizontal distance to the closest LoS or NLoS ABS:
\begin{align}
    &\begin{cases}
    \normalfont\text{I:}\,\,\,\, z\leq r_u-r_e \\
    \normalfont\text{II:}\,\,\,\, r_u-r_e<z\leq z_X(\beta^*) \\
    \normalfont\text{III:}\,\,\,\, z_X(\beta^*)<z\leq r_u+r_e\\
    \normalfont\text{IV:}\,\,\,\, z>r_u+r_e
    \end{cases}& \normalfont\text{if $r_u> r_e\,$,} \\
    &\text{and} \nonumber\\
    &\begin{cases}
    \normalfont\text{V:}\,\,\,\, z\leq r_e-r_u \\
    \normalfont\text{VI:}\,\,\,\, r_e-r_u<z\leq r_e+r_u \\
    \normalfont\text{VII:}\,\,\,\, z>r_e+r_u 
    \end{cases}& \normalfont\text{if $r_u\leq r_e\,$,}
\end{align}
where $\beta^*=\arcsin{\frac{r_e}{r_u}}$ and $z_X(\beta)=r_u\cos\beta+\sqrt{r_e^2-(r_u\sin\beta)^2}$ respectively denote the maximum angle $\beta$ and the maximum horizontal distance $z$ of the points describing the exclusion zone for any given angle $\pm\beta$. \par
The CDF of the horizontal projection of the distance between the user and the closest ABS in LoS or NLoS condition can be thus expressed as
\begin{align}
    F_{Z_M}(z)=\begin{cases} 
    \sum\limits_{i=I}^{I\!V} F_{Z_M}^i(z)\,\mathds{1}^i(z), & \normalfont\text{ if } r_u>r_e \nonumber\\
    \sum\limits_{i=V}^{V\!I\!I} F_{Z_M}^i(z)\,\mathds{1}^i(z), & \normalfont\text{ if } r_u\leq r_e   \end{cases},
\end{align}
where each indicator $\mathds{1}^i(z)$ is met if the circumference of radius $z$ falls into region $i$, and $F_{Z_M}^i(z)$'s are expressed in terms of $\kappa_M|_a^b=\int\limits_a^b \mathcal{P}_M({z})\,{z}\,{\rm d}{z}$ as follows:
\begin{align} 
\begin{cases}
F_{Z_M}^{I}(z)=&1-\exp\bigg(-2\pi\,\lambda_A\,\kappa_M|_0^z\bigg) \\
F_{Z_M}^{I\!I}(z)=&1-\exp\bigg(-\lambda_A\,\bigg(2\pi\,\kappa_M|_0^z-\int\limits_{-\beta_i(z)}^{\beta_i(z)}\kappa_M|_{z_m(\beta)}^z\,{\rm d}\beta\bigg)\bigg)\\
F_{Z_M}^{I\!I\!I}(z)=&1-\exp\bigg(-\lambda_A\bigg(2\pi\,\kappa_M|_0^z-\int\limits_{-\beta_i(z)}^{\beta_i(z)}\kappa_M|_{z_m(\beta)}^z\,{\rm d}\beta\nonumber\\
&-2\int\limits_{\beta_i(z)}^{\beta^*}\kappa_M|_{z_m(\beta)}^{z_X(\beta)}\,{\rm d}\beta\bigg)\bigg) \\
F_{Z_M}^{I\!V}(z)=&1-\exp\bigg(-\lambda_A\bigg(2\pi\,\kappa_M|_0^z-\int\limits_{-\beta^*}^{\beta^*}\kappa_M|_{z_m(\beta)}^{z_X(\beta)}\,{\rm d}\beta\bigg)\bigg)\\
\end{cases}, \label{eq:CDF_M1}\\
\end{align}
and
\begin{align}
\begin{cases}
F_{Z_M}^{V}(z)=&0 \\
F_{Z_M}^{V\!I}(z)=&1-\exp\bigg(-\lambda_A\bigg(2\pi\,\kappa_M|_0^z-\int\limits_{-\beta_i(z)}^{\beta_i(z)}\kappa_M|_0^z\,{\rm d}\beta\nonumber\\
&-\int\limits_{\beta_i(z)}^{2\pi-\beta_i(z)}\kappa_M|_0^{z_X(\beta)}\,{\rm d}\beta\bigg)\bigg)\\
F_{Z_M}^{V\!I\!I}(z)=&1-\exp\bigg(-\lambda_A\bigg(2\pi\,\kappa_M|_0^z-\int\limits_{-\pi}^{\pi}\,\kappa_M|_{0}^{z_X(\beta)} \,{\rm d}\beta\bigg)\bigg)
\end{cases}, \label{eq:CDF_M2}\\
\end{align}
where $\beta_i(z)=\arccos{\frac{z^2+r_u^2-r_e^2}{2\,z\,r_u}}$ is the angle of intersection between the circumference of radius $z$ centered at the user and the exclusion zone and $z_m(\beta)=r_u\cos\beta-\sqrt{r_e^2-(r_u\sin\beta)^2}$ is the minimum distance of any point inside the exclusion zone which angular coordinate is $\pm\beta$ (see Fig. \ref{fig:regions}).
Note also that $F_{Z_M}^{V\!I\!I}(z)=F_{Z_M}^{I\!V}(z)\bigg|_{ \genfrac{}{}{0pt}{}{z_m(\beta)=0}{\beta^*=\pi} }$. 
\end{theorem}

\begin{corollary} \label{cor:PDF_M}
The PDF of the horizontal projection of the distance between the user and the closest $N$ or $L$ ABS is given by
\begin{align}
    f_{Z_M}(z)=\begin{cases} 
    \sum\limits_{i=I}^{I\!V} f_{Z_M}^i(z)\,\mathds{1}^i(z), & \normalfont\text{ if } r_u>r_e \nonumber\\
    \sum\limits_{i=V}^{V\!I\!I} f_{Z_M}^i(z)\,\mathds{1}^i(z), & \normalfont\text{ if } r_u\leq r_e   \end{cases},
\end{align}
where the functions $f_{Z_M}^i(z)$'s are the derivatives of the respective $F_{Z_M}^i(z)$'s, namely
\begin{align}
&\begin{cases}
f_{Z_M}^{I}(z)=&2\pi\,\lambda_A\,z\,\mathcal{P}_M(z)\,\bar{F}_{Z_M}^I(z)\nonumber\\
f_{Z_M}^{I\!I}(z)=&2\,\lambda_A\,z\,\mathcal{P}_M(z)\bigg(\pi-\left(\frac{{\rm d}\beta_i(z)}{{\rm d}z}+\beta_i(z)\right)\bigg)\,\bar{F}_{Z_M}^{I\!I}(z)\nonumber\\
f_{Z_M}^{I\!I\!I}(z)=&2\,\lambda_A\bigg(\,z\,\mathcal{P}_M(z)\left(\pi-\frac{{\rm d}\beta_i(z)}{{\rm d}z}+\beta_i(z)\right)\nonumber\\
&+\kappa_M\bigg|_{z_m(\beta_i(z))}^{z_X(\beta_i(z))}\,\frac{{\rm d}\beta_i(z)}{{\rm d}z}\bigg)\,\bar{F}_{Z_M}^{I\!I\!I}(z)\nonumber\\
f_{Z_M}^{I\!V}(z)=&2\pi\,\lambda_A\,z\,\mathcal{P}_M(z)\,\bar{F}_{Z_M}^{I\!V}(z)\nonumber\\
\end{cases}, 
\end{align}
and
\begin{align}
&\begin{cases}
f_{Z_M}^{V}(z)=&0\nonumber \\
f_{Z_M}^{V\!I}(z)=&2\,\lambda_A\bigg(z\,\mathcal{P}_M(z)\left(\pi-\frac{{\rm d}\beta_i(z)}{{\rm d}z}+\beta_i(z)\right)\nonumber\\
&+\kappa_M|_0^{z_X(\beta_i(z))}\,\frac{{\rm d}\beta_i(z)}{{\rm d}z}\bigg)\,\bar{F}_{Z_M}^{V\!I}(z)\nonumber\\
f_{Z_M}^{V\!I\!I}(z)=&2\pi\,\lambda_A\,z\,\mathcal{P}_M(z)\,\bar{F}_{Z_M}^{V\!I\!I}(z)
\end{cases},
\end{align}
in which $\frac{{\rm d}\beta_i(z)}{dz}=\frac{r_u^2-z^2-r_e^2}{z\,\sqrt{4z^2r_u^2-(z^2+r_u^2-r_e^2)^2}}$.
\end{corollary}
\begin{IEEEproof} By applying Leibniz integral rule and taking the derivative of the CDFs in (\ref{eq:CDF_M1}) and (\ref{eq:CDF_M2}), the final results follow. \end{IEEEproof}

\subsection{Association Probabilities}
The association probabilities are defined as the probabilities that the average power received from the closest BS of a specific type is greater than the powers received from the closest BSs of any other type. 
   \begin{table*}[t]
  \begin{center}
    \caption{Minimum Interferer Distances $\left(d_B^C(z)\right)$}
    \label{tab:Distances}
     \begin{tabular}{|c|c|c|c|c|} \hline
      \textbf{B $\backslash$ C} & \textbf{L} & \textbf{N} & \textbf{T} \\
      \hline
      \textbf{L} & $\sqrt{z^2+h^2}$ & $\begin{cases}
      \frac{\eta_N}{\eta_L}^\frac{1}{\alpha_N}\,(\sqrt{z^2+h^2})^\frac{\alpha_L}{\alpha_N}, \, \text{ if } \sqrt{z^2+h^2}>d_N^L(h) \nonumber \\
      h, \, \text{ otherwise}
      \end{cases}$
      & $\frac{\xi_T}{\xi_L}^\frac{1}{\alpha_T}\,(\sqrt{z^2+h^2})^\frac{\alpha_L}{\alpha_T}$ \\ \hline
      \textbf{N} &  $\frac{\eta_L}{\eta_N}^\frac{1}{\alpha_L}\,(\sqrt{z^2+h^2})^\frac{\alpha_N}{\alpha_L}$
      & $\sqrt{z^2+h^2}$ & $\frac{\xi_T}{\xi_N}^\frac{1}{\alpha_T}\,(\sqrt{z^2+h^2})^\frac{\alpha_N}{\alpha_T}$  \\ \hline
      \textbf{T} & $\begin{cases}
      \frac{\xi_L}{\xi_T}^\frac{1}{\alpha_L}\,z^\frac{\alpha_T}{\alpha_L}, \, \text{ if } z>d_L^T(h) \nonumber \\
      h, \, \text{ otherwise}
      \end{cases}$
      & $\begin{cases}
      \frac{\xi_N}{\xi_T}^\frac{1}{\alpha_N}\,z^\frac{\alpha_T}{\alpha_N}, \, \text{ if } z>d_N^T(h) \nonumber \\
      h, \, \text{ otherwise}
      \end{cases}$
      & $z$ \\ \hline
    \end{tabular}  
  \end{center} 
\end{table*}

\begin{theorem} \label{thm:assoc}
Assuming the polar coordinate system $(\beta,z)$ centered around the user, let $B,C\in\{N,L,T\}$ respectively refer to the type of tagged and interferer BSs, then $d_B^C(z)$ represents the minimum Euclidean distance of any interferer when the tagged BS is located at horizontal distance $z$, and $z_B^C(z)=\begin{cases}d_B^C(z),\,&\text{if }C=T \\
\sqrt{\left(d_B^C(z)\right)^2-h^2},\,&\text{otherwise}\end{cases}$ denotes the horizontal projection of $d_B^C(z)$.
Let $M\in\{L,N\}$, then the association probabilities can be expressed as
\begin{align}
\mathcal{A}_M=&\int\limits_{\max(0,r_e-r_u)}^\infty f_{Z_M}({z})\,a_M({z})\,{\rm d}{z}, \\
\mathcal{A}_T=&\int\limits_0^\infty f_{Z_T}(z) \, a_T(z) \, {\rm d}z=1-\mathcal{A}_L-\mathcal{A}_N,
\end{align}
where $a_L(z)=\bar{F}_{Z_T}\left(d_L^T(z)\right)\,\breve{F}_{Z_N}\left(z_L^N(z)\right)$ with $\breve{F}_{Z_N}(z)=
\begin{cases}
1, &\text{if $z<z_L^N(0)$} \\  
\bar{F}_{Z_N}(z), &\text{otherwise}
\end{cases}$, $a_N(z)=\bar{F}_{Z_T}\left(d_N^T(z)\right)\,\bar{F}_{Z_L}(z_N^L(z))$, and $a_T(z)=\bar{F}_{Z_L}\left(d_T^L(z)\right)\,\bar{F}_{Z_N}(d_T^N(z))$ respectively represent the association probabilities conditioned on the association to $L$, $N$, and $T$ BSs, which we refer to as conditional $L$-, $N$-, and $T$-association probabilities. 
   \end{theorem}
 \begin{IEEEproof}See Appendix \ref{appx:assoc}. \end{IEEEproof}
   
\subsection{Conditional Laplace Transform of the Interference}
\normalfont{Since all the base stations are assumed to work on the same frequency spectrum, they interfere with each other. 
By computing the Laplace transform of the random interference $I$ we can essentially characterize the interference statistics.
Each conditional Laplace transform of the interference is the product of the conditional Laplace transforms of each type of interferers.}

\begin{theorem}\label{thm:Lap_T}
Let $r(\beta',z')=\sqrt{r_u^2+z'^2-2\,r_u \, z' \,\cos\beta'}$ denote the horizontal distance between a generic TBS located at $(\beta',z')$ and the origin.
The conditional Laplace transform of the interference coming from the interfering TBSs when the serving BS is located at distance $z$ can be expressed as 
\begin{align} \label{eq:Lap_T}
\mathcal{L}_{I,B}^T(s,z)=&\exp\Bigg(-\lambda_T\int\limits_{z_B^T(z)}^\infty\int\limits_0^{2\pi}
\bigg(1-\left(\frac{m_T}{m_T+s\,\xi_T\,{z'}^{-\alpha_T}}\right)^{m_T}\bigg)\nonumber\\
&\times\mathcal{G}_T\left(r(\beta',{z'})\right)\,{z'}\,{\rm d}\beta'\,{\rm d}{z'}\Bigg)\,,
\end{align}
where $B$ refers to the type of the tagged BS and $\mathcal{G}_T(r)$ is the PDF of the distribution of the TBSs' distribution with respect to the origin.
\end{theorem}
\begin{IEEEproof}See Appendix \ref{appx:Lap_T}. \end{IEEEproof}

\begin{theorem}\label{thm:Lap_M}
Let $\mathds{1}^i$ be defined as in Theorem \ref{thm:CDF_M}, then the Laplace transform of the interference coming from either the LoS or the NLoS ABSs can be expressed as 
\begin{align} 
\mathcal{L}_{I,B}^M(s,z)= 
\begin{cases}
    \sum\limits_{i=I}^{I\!V}\mathcal{L}_{I,B}^{M,i}(s,z)\,\mathds{1}^i(z_B^M(z)), & \normalfont\text{ if $r_u>r_e$} \\
    \sum\limits_{i=V}^{V\!I\!I}\mathcal{L}_{I,B}^{M,i}(s,z)\,\mathds{1}^i(z_B^M(z)), & \normalfont\text{ if $r_u\leq r_e$}
\end{cases},
\end{align}
where
\begin{align}
\begin{cases}
\mathcal{L}_{I,B}^{M,I\!V}(s,z)=&\exp\big(-2\pi\lambda_A \int\limits_{z_B^M(z)}^\infty \mathcal{I}_M(s,z')\,{\rm d}z'\big) \\
\mathcal{L}_{I,B}^{M,I\!I\!I}(s,z)=&\mathcal{L}_{I,B}^{M,I\!V}(s,z)\nonumber\\
&\times\exp\big(\lambda_A \int\limits_{-\beta_i({z_B^M(z)})}^{\beta_i({z_B^M(z)})}\int\limits_{z_B^M(z)}^{z_X(\beta)} \mathcal{I}_M(s,z')\,{\rm d}z'\,{\rm d}\beta'\big) \\
\mathcal{L}_{I,B}^{M,I\!I}(s,z)=&\mathcal{L}_{I,B}^{M,I\!I\!I}(s,z)\nonumber\\
&\times\exp\big(2\lambda_A \int\limits_{\beta_i({z_B^M(z)})}^{\beta^*}\int\limits_{z_B^M(z)}^{z_X(\beta)} \mathcal{I}_M(s,z')\,{\rm d}z'\,{\rm d}\beta'\big) \nonumber\\
\mathcal{L}_{I,B}^{M,I}(s,z)=&\mathcal{L}_{I,B}^{M,I\!V}(s,z)\nonumber\\
&\times\exp\big(\lambda_A \int\limits_{-\beta^*}^{\beta^*}\int\limits_{z_m(\beta)}^{z_X(\beta)} \mathcal{I}_M(s,z')\,{\rm d}z'\,{\rm d}\beta'\big) 
\end{cases}, \end{align}
and
\begin{align} \begin{cases}
\mathcal{L}_{I,B}^{M,V\!I\!I}(s,z)=&\exp\big(-2\pi\lambda_A \int\limits_{z_B^M(z)}^\infty \mathcal{I}_M(s,z')\,{\rm d}z'\big) \nonumber\\
\mathcal{L}_{I,B}^{M,V\!I}(s,z)=&\mathcal{L}_{I,B}^{M,V\!I\!I}(s,z)\nonumber\\
&\times\exp\big(\lambda_A \int\limits_{-\beta_i(z_B^M(z))}^{\beta_i(z_B^M(z))}\int\limits_{z_B^M(z)}^{z_X(\beta)} \mathcal{I}_M(s,z')\,{\rm d}z'\,{\rm d}\beta'\big) \nonumber\\
\mathcal{L}_{I,B}^{M,V}(s,z)=&\mathcal{L}_{I,B}^{M,V\!I\!I}(s,z)\nonumber\\
&\times\exp\big(\lambda_A \int\limits_{-\pi}^{\pi}\int\limits_{z_B^M(z)}^{z_X(\beta)} \mathcal{I}_M(s,z')\,{\rm d}z'\,{\rm d}\beta'\big)\,, 
\end{cases} \end{align}
in which the function $\mathcal{I}_M(s,z')=\bigg(1-\Big(\frac{m_M}{m_M+s\,\xi_M\,\left(d_M^M(z')\right)^{-\alpha_M}}\Big)^{m_M}\bigg)\,z'\,\mathcal{P}_M(z')$ has been introduced.
Note also that $\mathcal{L}_{I,B}^{M,V\!I\!I}(s,z)=\mathcal{L}_{I,B}^{M,I\!V}(s,z)$.
\end{theorem}
\begin{IEEEproof}See Appendix \ref{appx:Lap_M}. \end{IEEEproof}

\begin{corollary}
Let $C\in\{L,N,T\}$, then the conditional Laplace transform can be expressed as 
\begin{align}
\mathcal{L}_{I,B}(s,z)=&\mathcal{L}_{I,B}^{L}(s,z)\,\mathcal{L}_{I,B}^{N}(s,z)\,\mathcal{L}_{I,B}^{T}(s,z)\nonumber\\=&\prod_{C}\mathcal{L}_{I,B}^C(s,z)\,.
\end{align}
\end{corollary}

\subsection{Coverage Probability}
Based on the expressions derived for the PDFs of the distance to the closest BS, the conditional association probabilities, the SINR, and the Laplace transform of the interference, hereby we finally provide the exact and the approximate expressions of the coverage probability.

\begin{theorem}\label{thm:Pc}
Let $P_{c,B}$ denote the exact coverage probability conditioned on the association to a $B=\{L,N,T\}$ BS, then the exact coverage probability for a typical user in the VHetNet described in Section \ref{sec:S} is given by
\begin{align}
     {P}_c=&\int\limits_0^\infty a_T({z}) \, P_{c,T}({z}) \, f_{Z_T}({z})\,{\rm d}{z} \nonumber\\
&+\sum\limits_{M=L,N}\int\limits_{\max(0,r_e-r_u)}^\infty a_M\Big(\sqrt{{z}^2+h^2}\Big) \nonumber\\
&\times P_{c,M}\left(\sqrt{{z}^2+h^2}\right) \, f_{Z_M}({z})\,{\rm d}{z},
\end{align}
where $P_{c,B}=\E_{Z_B}\bigg[\sum\limits_{k=0}^{m_B-1} \bigg(\frac{(-s)^k}{k!}\,\frac{\partial^k}{\partial s^k} \mathcal{L}_{J,B}(s,z)\bigg)_{s=\mu_B(Z_B)}\bigg]$ with $Z_B$ denoting the random horizontal distance from the tagged BS, and $J=\sigma_n^2+I$.
The expressions of the functions $f_{Z_B}(z)$'s, $a_B(z)$'s, and $\mathcal{L}_{I,B}(s,z)$'s are provided in Corollaries \ref{cor:PDF_T} and \ref{cor:PDF_M}, Theorem \ref{thm:assoc}, and Theorems \ref{thm:Lap_T} and \ref{thm:Lap_M}, respectively.
\end{theorem}
\begin{IEEEproof}
  See Appendix \ref{appx:Pc}.
\end{IEEEproof}

Since computing the exact expression of the coverage probability may require high order derivatives of the Laplace transform of the interference, it is usually more convenient to make use of the approximations suggested by the following theorem.

\begin{theorem}\label{thm:Pc_tilde}
To ease the evaluation of the exact coverage probability introduced in Theorem \ref{thm:Pc}, the conditional coverage probability can be approximated as
\begin{align} \label{eq:approx_PcB}
\tilde{P}_{c,B}=\sum\limits_{k=1}^{m_B}
\binom{m_B}{k}\,(-1)^{k+1}\,\mathcal{L}_{J,B}\left(k\,\varepsilon_{2,B}\,\nu_B,z\right),
\end{align}
where $\varepsilon_{2,B}=(m_B!)^{-\frac{1}{m_B}}\,$ and $\nu_B=m_B\,\frac{\tau}{\xi_B}\,\left(d_B^B(z)\right)^{\alpha_B}$.
\end{theorem}
\begin{IEEEproof}See Appendix \ref{appx:Pc_tilde}. \end{IEEEproof}
  
\section{Results and Discussion} \label{sec:NR}
Hereby we present the analytical results based on the previously derived expressions and validate them by means of simulations.
The results we obtained explain the influence of various system parameters on the overall performance of the network, where TBSs are deployed according to a 2D Gaussian distribution, and hence $\mathcal{G}_T(r)=\frac{1}{\sigma_T\,\sqrt{2\,\pi}}\,e^{-\frac{r^2}{2\,\sigma_T^2}},\,\forall r\geq0$. 
In particular, we have chosen $\lambda_T$ and $\sigma_T$ in order to have more than 8 TBSs/km$^2$ within a distance of 2 km from the town center (urban environment) and less than 0.1 TBSs/km$^2$ when the distance exceeds 10 km (rural environment). \par
Whenever omitted, the values of the system parameters mentioned in this section refer to table \ref{tab:Parameters}. 
 \begin{table}[t] 
  \begin{center}
    \caption{Main System Parameters Used in Simulations}
    \label{tab:Parameters}
    \begin{tabular}{|c|c|} 
    \hline
      \textbf{Parameters} &
      \textbf{Values} \\\hline
      Mean additional transmit losses &
       $\begin{cases}
       \eta_L=-0.1\,\text{dB}=0.9772 \\
       \eta_N=-21\,\text{dB}=0.007943 \\
       \eta_T=-1.6\,\text{dB}=0.6918
       \end{cases}$  \\\hline
       Path loss exponents &
       $\begin{cases}
       \alpha_L=3 \\
       \alpha_N=4 \\
       \alpha_T=3.5
       \end{cases}$ \\
       \hline
       Nakagami shape parameters &
       $\begin{cases}
       m_L=2 \\
       m_N=1 \\
       m_T=1
       \end{cases}$ \\\hline
       BSs' densities & 
       $\begin{cases}\lambda_A=0.15\text{ ABSs/km$^2$}\\ \lambda_T=8\times10^4\text{ TBSs/km$^2$}\end{cases}$\\
       \hline
       Transmission powers &
     $\begin{cases}p_M=2\,\text{dB}=1.585\text{ W}\\
     p_T=10\,\text{dB}=10\text{ W} \end{cases}$ \\\hline
     S-curve parameters &
       $\begin{cases} \mathcal{S}_a=4.88\\
       \mathcal{S}_b=0.429 \end{cases}$ \\ 
    \hline
    Variance of TBSs' distribution  &
     $\sigma_T^2=10\,\text{km}^2$ \\\hline 
     SINR threshold &
     $\tau=-5\,\text{dB}=0.3162$\\ \hline
     Noise power spectral density &
     $\sigma_n^2=-120\,\text{dB}=10^{-12}\,\frac{\text{W}}{\text{Hz}}$ \\\hline
     ABSs' altitude &
        $h=100\,\text{m}$ \\ 
     \hline
     \end{tabular}  
  \end{center}
\end{table} 

\subsection{Coverage Probability}
\begin{figure}[t]
\centering
\includegraphics[width=1\columnwidth]{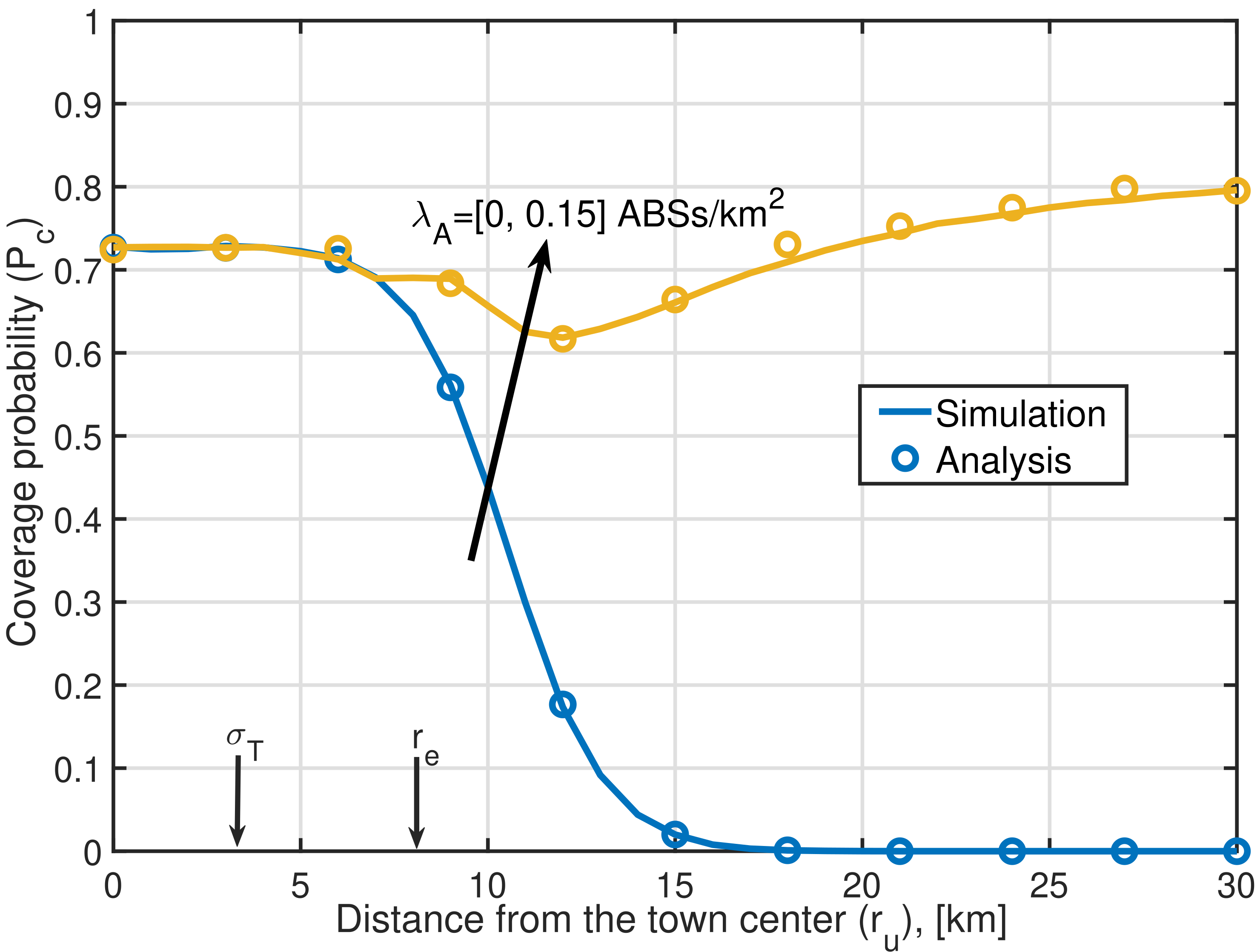}
\caption{Numerical results for the coverage probability as function of the distance of the user from the town center.
The analytical results have been obtained by computing the exact coverage probability when $\lambda_A=0$ and the approximate coverage probability when $\lambda_A=0.15\,$ABSs/km$^2$.}
\label{fig:Pc}
\end{figure}

\begin{figure}[t]
\centering
\includegraphics[width=1\columnwidth]{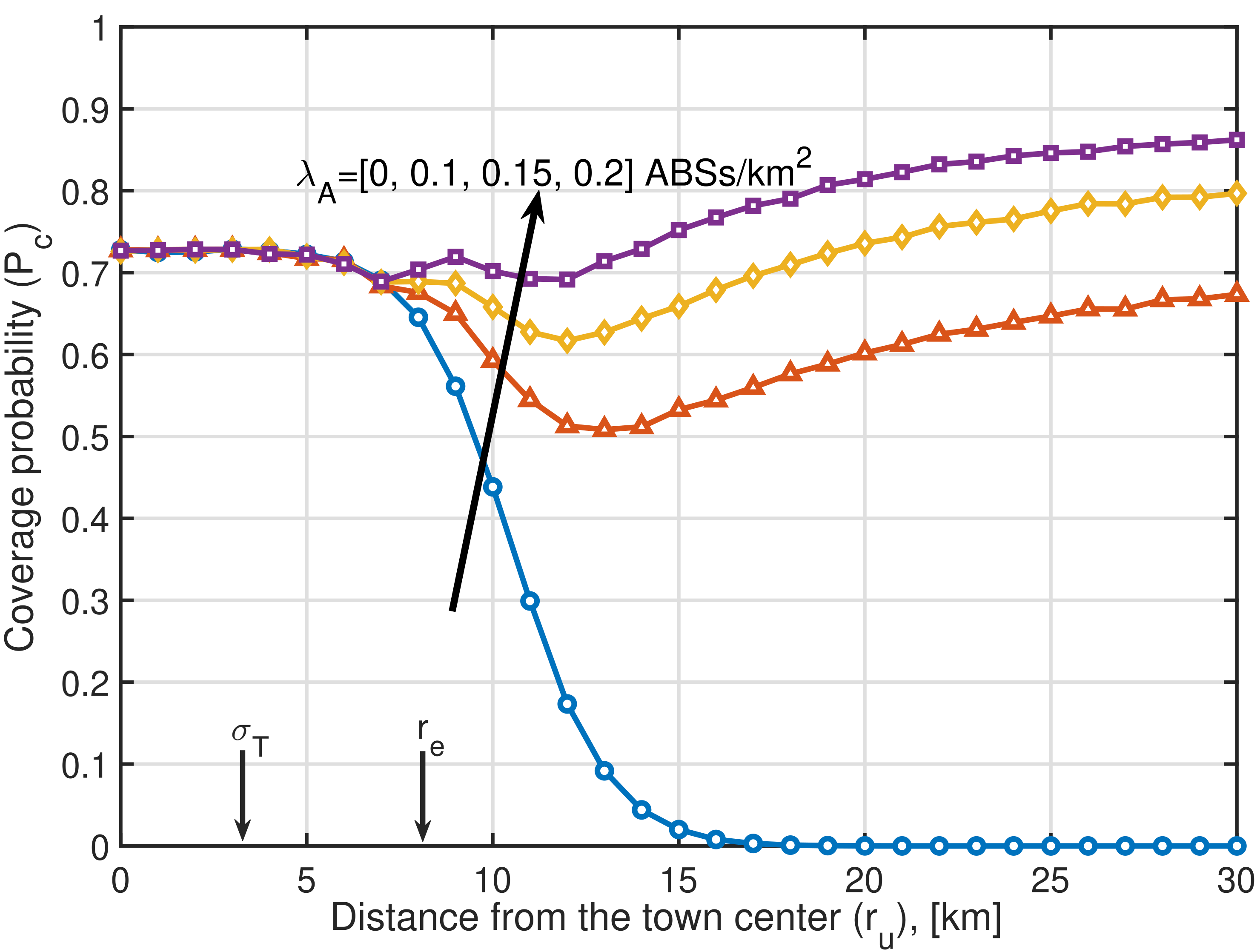}
\caption{Simulation results for the coverage probability as function of the distance of the user from the town center with various ABSs densities.}
\label{fig:Pc_s}
\end{figure}

To prove the correctness of our mathematical model, the behaviour of the coverage probability versus the distance $r_u$ is shown Fig. \ref{fig:Pc}, where the analytical results show the exact coverage probability for the terrestrial network and the approximate coverage probability for the UAV-assisted one.
The influence of the density of ABSs on the coverage probability is highlighted in Fig. \ref{fig:Pc_s}.
When no UAVs are deployed, the coverage probability generally decreases as moving away from the town center, and hence rural users are rarely served.
On the other hand, assuming $r_e=8\,$km, deploying drones with a density of 0.15$\,$ABSs/km$^2$ allows to even the coverage probability with negligible deterioration for urban users, and by going up to just 0.2 ABSs/km$^2$ rural zones become much better covered than urban ones (see Fig. \ref{fig:Pc_s}). 
This is because the UAVs are deployed only in the periphery and hence the urban users experience interference from almost only far $N$ ABSs, receiving a negligible power from them.
On the other hand, rural users can see many more $L$ ABSs and, since TBSs are sparse, mostly associate with one of them taking advantage of a better communication channel (see Fig. \ref{fig:A_LT}) and hence the coverage probability is consistently improved.

\subsection{Association Probabilities}
\begin{figure}[t]
\centering
\includegraphics[width=1\columnwidth]{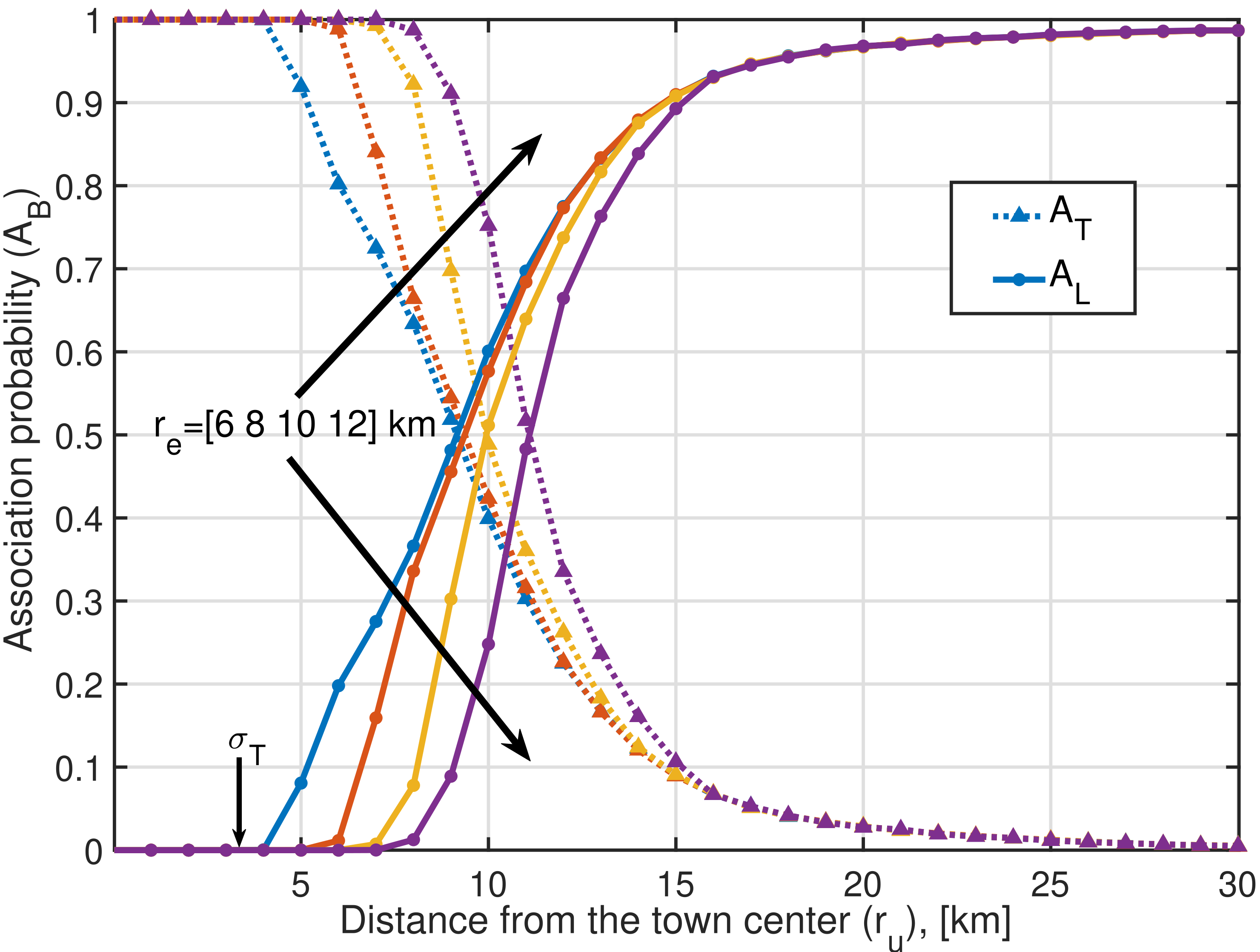}
\caption{Simulation results for the probability of association to LoS ABSs and TBSs (the ones for NLoS ABSs are the complement to unity, and are always negligible) as function of the distance of the user from the town center for various exclusion radii. 
The results have been obtained by averaging over $100\,000$ iterations.}
\label{fig:A_LT}
\end{figure}

To better understand the results previously obtained, Fig. \ref{fig:A_LT} shows the behaviors of the association probabilities to $L$ and $T$ BSs\footnote{The association probability to $N$ BSs has not been plotted since it is the complement to unity of the other two association probabilities, and it is negligible over the whole range of $r_u$ considered.}.
It is evident that when deploying UAVs and fixing a value of $r_e$, three regimes can be identified: for $r_e=8$ km, the first regime is characterized by associations to TBSs only and it occurs within $0$-$7$ km, whereas the second and most unpredictable regime is a mix of $T$ and $L$ associations and the third one is mainly characterized by $L$ associations and a small percentage of $T$ and $N$ ones, starting from around 25 km. 
As the exclusion zone expands, the first regime is also extended to larger distances from the town center, but the transition to the second regime becomes sharper because of the low density of TBSs. 
Intuitively, the exclusion radius has almost no influence on the associations in the third regime since they happen at much larger values of $r_u$, where the environment has become quite homogeneous.\par
Revisiting Fig. \ref{fig:Pc_s}, we observe that when UAVs are deployed the coverage probability experiences a local minimum at $r_u=11$-$13\,$km (second regime).
Within this interval, indeed, there is an excessive drop in coverage probability whenever the user associates to a TBS, as shown by the blue curve for $\lambda_A=0$, and therefore the required density of UAVs is maximum.
This is definitely a nontrivial result since the support provided by LoS UAVs already starts being evident at the edge of the exclusion zone, where the association to ABSs is a relatively rare event ($\mathcal{A}_L$ equals just one third at $r_u=r_e=8\,$km). 

\subsection{Minimum Coverage Probability}
\begin{figure}[t]
\centering
\includegraphics[width=1\columnwidth]{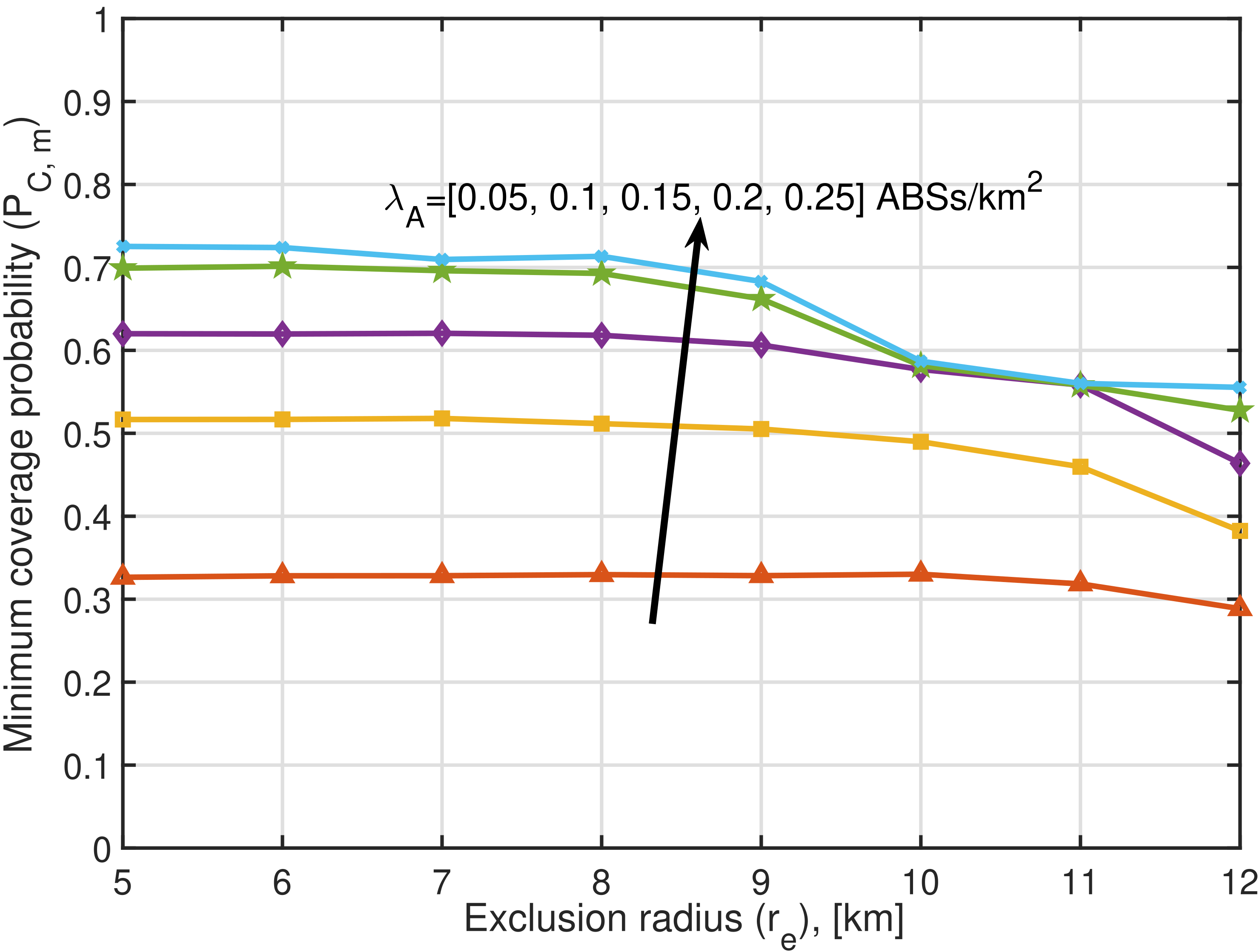}
\caption{Simulation results for the minimum coverage probabilities (among the values of $r_u$ from 0 to 30 km with step 3 km) versus the exclusion radius with various ABSs densities.}
\label{fig:Pc_m}
\end{figure}

Let the minimum coverage probability ($P_{c,m}$) be defined as the minimum value of the coverage probability within the range of values of $r_u$ considered in Fig. \ref{fig:Pc_s}, for a given exclusion radius and a given density of ABSs.
Fig. \ref{fig:Pc_m} shows that increasing the density of UAVs generally ameliorates $P_{c,m}$, especially for relatively small values of $r_e$.
However, it is expected for the value of $P_{c,m}$ to start decreasing when $\lambda_A$ exceeds a critical value, since eventually the effect of increasing the aerial interference would dominate over the effect of reducing the average minimum distance to the serving BS.
However, this critical value is larger than those simulated in Fig. \ref{fig:Pc_m}. \par
\begin{figure}[t]
\centering
\includegraphics[width=1\columnwidth]{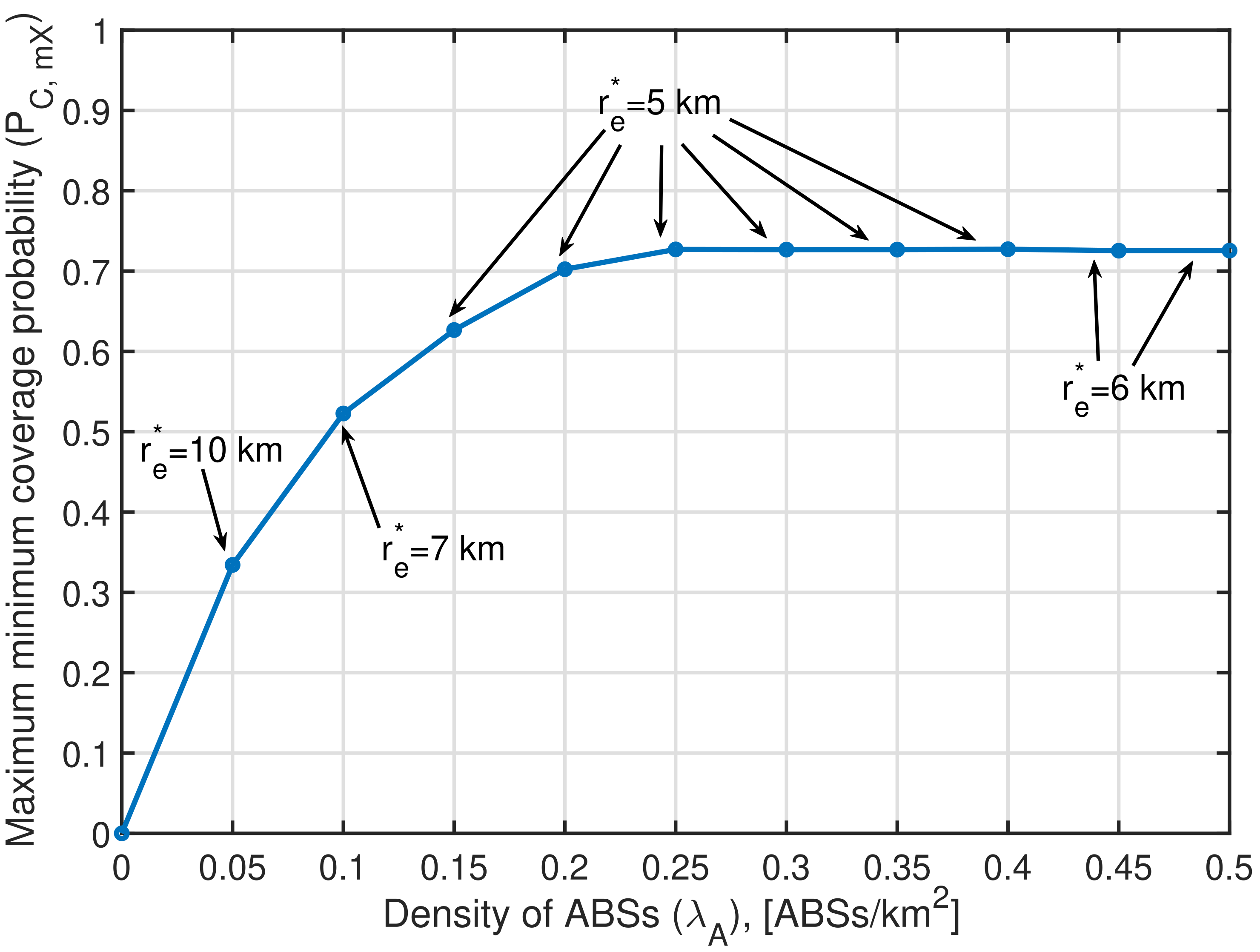}
\caption{Simulation results for the maximum of the minimum coverage probability as a function of the ABS density. 
This plot has been obtained by replicating the simulations in Fig. \ref{fig:Pc_m} for many values of $\lambda_A$ and extracting the maximum value of the minimum coverage probability for each.
The labels indicate the values of $r_e^*$, which is defined as the largest value of $r_e$ that allows to achieve the minimum coverage probability given a specific density of ABSs.}
\label{fig:maxPc_m}
\end{figure}

Another interesting result is shown in Fig. \ref{fig:maxPc_m}, where the maximum (over the values of $r_u$ going from 0 to 30 km with step 3 km) of the minimum coverage probability is plotted versus the density of deployed UAVs considering our standard parameters.
The figure shows that this quantity rapidly saturates over $\lambda_A\approx0.2$ ABSs/km$^2$, never exceeding 74$\%$.
The best scenario expects to start deploying UAVs at distances greater or equal to $r_e^*$, that is the maximum exclusion zone for which the maximum $P_{c,m}$ is achieved. 
However, for $\lambda_A=0.15$ ABSs/km$^2$ it is evident from Fig. \ref{fig:Pc_m} that the benefit in terms of minimum coverage probability would be negligible compared to the case with $r_e=8$ km, and probably not convenient since it would require to deploy a surplus of around eighteen UAVs. 
For very small values of $\lambda_A$, $r_e^*$ is large because the interference coming from TBSs dominates at any value of $r_u$, and hence, it is better not to deploy UAVs close to the town center.
The value of $r_e^*$ keeps decreasing until a point of minimum around $\lambda_A=0.2$, over which the density of ABSs is such that aerial interference becomes predominant for $r_u>r_e$, and thus it becomes convenient to extend the exclusion zone.

\section{Conclusion and Future Work} \label{sec:C}
A new stochastic geometry framework has been presented in this paper in order to evaluate the coverage probability in a typical region including urban and rural areas, where TBSs are typically getting sparse as we move away from the city center and ABSs are supposed to compensate this deficiency. 
Thus, ABSs have been introduced and analytical expressions have been derived for the distribution of the distances between a typical user and the tagged BS as well as the association probabilities. 
Exact and approximate expressions for the overall coverage probability were derived as a function of the distance between the user and the town center.
We showed that supporting the terrestrial infrastructure with ABSs in peripheral areas results in an overall improvement of the quality of service (QoS), although the number of interferers consequently grows. 
This actually means that the users close to the urban center will slightly lose coverage, but the benefit for all the remaining users definitely outperforms this drawback.\par
This work can be extended in multiple directions. 
For instance, different types of aerial platforms can be considered such as HAPs or tethered balloons, depending on the ground users' spatial distribution and the requirements of the network.
Furthermore, another interesting extension to this work would consider the use of TV white space (TVWS) spectrum band for aerial platforms, given that TVWS is well-known to have a good potential to enhance coverage in rural areas.
   
\appendices
\section{Proof of Theorem \ref{thm:CDF_T}}\label{appx:CDF_T}
Recalling that the altitude of the TBS antennas is assumed to be negligible compared to the altitude of the ABSs, let $Z_T$ indicate either the Euclidean distance between the user and the tagged TBS or the horizontal projection of the same distance, then the expression of the respective CDF can be derived from the null probability of the PPP \cite{Andrews11}. 
For simplicity, we consider the polar coordinate system $(\beta,z)$.
\begin{align}    
F_{Z_T}(z)=&\P(Z_T\leq z)=1-\P(Z_T>z)\nonumber\\
=&1-\P[N_T(r_u,z)=0]
=1-\exp \bigg({-\int\limits_{\mathscr{B}_0}\tilde{\lambda}_T(r) \,{\rm d}S}\bigg) \nonumber\\
=&1-\exp\bigg(-\lambda_T\int\limits_0^z\int\limits_{-\pi}^\pi \mathcal{G}_T(r(\beta,z'))\,r(\beta,z')\,{\rm d}\beta\,{\rm d}z'\bigg),
\end{align} 
where $N_T(r_u,z)$ denotes the number of TBSs located inside a circle of radius $z$ centered around the typical user, $\mathscr{B}_0$ describes the ball of radius $z$ centered around the user, and $r(\beta, z)=\sqrt{r_u^2+z^2-2\,z\,r_u\,\cos\beta}$.

\section{Proof of Theorem \ref{thm:assoc}} \label{appx:assoc}
The association probability is defined as the probability that the average power received from the closest BS of a specific type is greater than the power received from the closest BSs of any other type. \par
Let $d_B^C(z)$ be the minimum Euclidean distance of any interfering BS of type $C$, given that the user associates to a BS of type $B$ located at horizontal distance $z$, since the corresponding Euclidean distance equals $z$ if $B=T$ and $\sqrt{z^2+h^2}$ otherwise, it follows that $z_B^C(z)=\begin{cases}
d_B^C(z),\,&\text{if }C=T \\
\sqrt{\left(d_B^C(z)\right)^2-h^2},\,&\text{otherwise}\end{cases}$ defines the horizontal projection of $d_B^C(z)$. \\
Recalling that  $\xi_Q=p_Q\,\eta_Q$ and introducing the random Euclidean distance $D_Q=d_Q^Q(Z_Q)$, the association probabilities will be derived for each type of BSs.

\subsubsection{Association Probability for $L$ BSs}
For this association probability we have to consider that there is a distance $d_N^L(0)$ below which the user surely associates with a LoS; indeed, since $\eta_L>\eta_N$, $\alpha_L<\alpha_N$, and all the ABSs are assumed to be deployed at the same altitude, we have
\begin{align}
\P\bigg(D_L<\left(\frac{\eta_L}{\eta_N}\right)^{\frac{1}{\alpha_L}}\,D_N^{\frac{\alpha_N}{\alpha_L}}\bigg)=&1, \nonumber\\
& \normalfont\text{ if $h<D_L<d_N^L(0)$}, 
\end{align}
where $d_N^L(0)=\left(\frac{\eta_L}{\eta_N}\right)^{\frac{1}{\alpha_L}}\,h^{\frac{\alpha_N}{\alpha_L}}$. This implies
\begin{align}
d_L^N(Z_L)=\begin{cases} 
h, & \normalfont\text{ if $h<D_L<d_N^L(0)$} \\
\left(\frac{\eta_N}{\eta_L}\right)^\frac{1}{\alpha_N}\,D_L^{\frac{\alpha_L}{\alpha_N}}, &\normalfont\text{ if $D_L>d_N^L(0)$} 
\end{cases}.\end{align}

Since the probabilities that the type of the tagged BS is $L$ rather than $N$ and $L$ rather than $T$ both depend on ${D}_L$, then
\begin{align}
\mathcal{A}_L=&\mathbb{E}_{Z_L}\big[\mathds{1}(Z_L\leq z_N^L(0))\,\P(Z_T>d_L^T(Z_L)\,|\,Z_L=z)\nonumber\\
&+\mathds{1}(Z_L>z_N^L(0))\,\P(Z_T>d_L^T(Z_L)\,|\,Z_L=z)\nonumber\\
&\times\P(Z_N>z_L^N(Z_L)\,|\,Z_L=z)\big] \nonumber \\ 
=& \int\limits_{\max(0,r_e-r_u)}^\infty f_{Z_L}(z)\,\bar{F}_{Z_T}\big(d_L^T(z)\big)\,\breve{F}_{Z_N}\big(z_L^N(z)\big)\,{\rm d}{z} \nonumber\\
=&\int\limits_{\max(0,r_e-r_u)}^\infty f_{Z_L}(z)\,a_L(z)\,{\rm d}{z},
\end{align}
where the function $\breve{F}(z)=
\begin{cases}
1, &\text{if $z<d_L^N(0)$} \\  
\bar{F}(z), &\text{otherwise}
\end{cases}$ has been introduced in order to compute the conditional $L$-association probability $a_L(z)=\bar{F}_{Z_T}\big(d_L^T(z)\big)\,\breve{F}_{Z_N}\big(z_L^N(z)\big)$. 

\subsubsection{Association Probability for $N$ BSs} 
Let us firstly recall that $d_N^T(Z_N)=\left(\frac{\xi_T}{\xi_N}\right)^{\frac{1}{\alpha_T}}\,Z_N^{\frac{\alpha_N}{\alpha_T}}$ and $d_N^L(Z_N)=\left(\frac{\xi_L}{\xi_N}\right)^{\frac{1}{\alpha_L}}\,Z_N^{\frac{\alpha_N}{\alpha_L}}=\left(\frac{\eta_L}{\eta_N}\right)^{\frac{1}{\alpha_L}}\,Z_N^{\frac{\alpha_N}{\alpha_L}}$, where the last equality is justified by the fact that all ABSs have the same transmission power $p_M$. \par
Now, the association probability for $N$ BSs can be obtained as
\begin{align}
\mathcal{A}_N=&\mathbb{E}_{R_N}\big[\P\left(Z_T>d_L^T(Z_N)\,|\,Z_N\right)\,\P\left(Z_L>z_L^N(Z_N)\,|\,Z_N\right)\big]     \nonumber\\
=&\int\limits_{\max(0,r_e-r_u)}^\infty{f_{Z_N}(z)\,\bar{F}_{Z_T}\left(d_N^T(z)\right)\,\bar{F}_{Z_L}\left(z_N^L(z)\right)\,{\rm d}z} \\
=&\int\limits_{\max(0,r_e-r_u)}^\infty{f_{Z_N}(z)\,a_N(z)\,{\rm d}z},  
\end{align}
where $a_N(z)=\bar{F}_{Z_T}\left(d_N^T(z)\right)\,\bar{F}_{Z_L}\left(z_N^L(z)\right)$ is the conditional $N$-association probability.

\subsubsection{Association Probability for $T$ BSs} In this last case, the association probability can be obtained as the complement to unity of the sum of the other two association probabilities, that is 
$$\mathcal{A}_T=1-\mathcal{A}_L-\mathcal{A}_N,$$
   or, alternatively, it can be computed as
   \begin{align}
   \mathcal{A}_T=&\int\limits_0^\infty f_{Z_T}(z) \, \bar{F}_{Z_L}\left(d_T^L(z)\right)\,\bar{F}_{Z_N}\left(d_T^N(z)\right)\,{\rm d}z\nonumber\\
   =& \int\limits_0^\infty f_{Z_T}(z) \, a_T(z) \, {\rm d}z,
   \end{align}
where $a_T(z)=\bar{F}_{Z_L}\left(d_T^L(z)\right)\,\bar{F}_{Z_N}\left(d_T^N(z)\right)$ denotes the conditional $T$-association probability.
   
\section{Proof of Theorem \ref{thm:Lap_T}} \label{appx:Lap_T}
Assuming a polar coordinate system centered around the user, the horizontal distance $r$ from the origin to any TBS located at $(\beta',z')$ can be expressed as $r(\beta',z')=\sqrt{r_u^2+z'^2-2\,r_u \, z' \,\cos\beta'}$. 
To obtain the expression of the conditional Laplace transform of the interference generated by TBSs, firstly we have to take the expectation over both the point process and the set of fading gains \cite{PRIMER}:
\begin{align}
\mathcal{L}_{I,B}^T(s,z)=&\E\left[e^{-s\,I_T}\right]\nonumber\\
\overset{(a)}{=}&\E_{\Phi_T} \bigg[\prod_{Y_i\in\Phi_T \backslash\{\mathscr{B}_0\}} \psi_T(s,Y_i)\bigg]\nonumber\\
\overset{(b)}{=}&\exp\bigg(-\int\limits_{\R^2 \backslash \{\mathscr{B}_0\}} \tilde{\lambda}_T(X)\,(1-\psi_T(s,X))\,{\rm d}X\bigg) \nonumber \\
=&\exp\bigg(-\lambda_T\int\limits_{z_B^T(z)}^\infty\int\limits_0^{2\pi}\mathcal{G}_T(r(\beta',z'))\nonumber\\
&\times\bigg(1-\Big(\frac{m_T}{m_T+s\,\xi_T\,z'^{-\alpha_T}}\Big)^{m_T}\bigg)\,z'\,{\rm d}\beta'\,{\rm d}z'\bigg),
\end{align}
 where, (a) follows from the independence of the exponentially distributed gains $G_{T,Y_i}$'s, having introduced the function $\psi_C(s,W_i)=\E_{G_C}\left[\exp\left(-\frac{s\,G_{C,W_i}\,p_C}{\|W_i\|^{\alpha_C}}\right)\right]$ for any type of interferers, and (b) derives from the application of the probability generating functional (PGFL) to the latter function.

\section{Proof of Theorem \ref{thm:Lap_M}} \label{appx:Lap_M}
Having obtained the general expression for the conditional Laplace transform of the terrestrial interference, the aerial one can be easily derived by simply identifying the domains of the interfering ABSs and applying the same procedure shown in Appendix \ref{appx:Lap_T}.
Each domain essentially is the projection at altitude $h$ of the complement\footnote{For any given set $S$, we identify its complement as $S'$.} of the circle of radius $z_B^M(z)$ centered around the user.
According to Fig. \ref{fig:regions}, for each of the regions hosting the respective circumference of radius $z_B^M(z)$, we can easily identify the resulting domain of the aerial interferers as one of the following:
\begin{align}&\begin{cases}
\normalfont\text{IV}'=&\left[z_B^M(z),\infty\right] \nonumber\\
\normalfont\text{III}'=&\left[z_B^M(z),\infty\right]\nonumber\\
&-\left[-\beta_i(z_B^M(z)),\beta_i(z_B^M(z))\right]\times\left[z_B^M(z),z_X(\beta)\right]\nonumber\\
\normalfont\text{II}'=&\left[z_B^M(z),\infty\right]\nonumber\\
&-\left(\left[-\beta_i(z_B^M(z)),-\beta^*\right]\cup\left[\beta_i(z_B^M(z)),\beta^*\right]\right)\nonumber\\
&\,\times\left[z_B^M(z),z_X(\beta)\right]\nonumber\\
\normalfont\text{I}'=&\left[z_B^M(z),\infty\right]-\left[-\beta^*,\beta^*\right]\times\left[z_m(\beta),z_X(\beta)\right] \nonumber\\
 \end{cases}, \end{align}
if $r_u>r_e$, and
\begin{align}
&\begin{cases}
\normalfont\text{VII}'=&\left[z_B^M(z),\infty\right] \nonumber\\
\normalfont\text{VI}'=&\left[z_B^M(z),\infty\right]\nonumber\\
&-\left[-\beta_i(z_B^M(z)),\beta_i(z_B^M(z))\right]\times\left[z_B^M(z),z_X(\beta)\right]\nonumber\\
\normalfont\text{V}'=&\left[z_B^M(z),\infty\right]-\times\left[z_B^M(z),z_X(\beta)\right]
 \end{cases}, \end{align}
if $r_u\leq r_e$, which indeed are consistent with the respective integration limits in the expressions of the various $\mathcal{L}_{I,B}^{M,i}(s,z)$'s.

\section{Proof of Theorem \ref{thm:Pc}} \label{appx:Pc}
Recalling from table \ref{tab:Distances} that $d_B^B(z)=\begin{cases}
z, &\text{ if } B=T \nonumber\\ 
\sqrt{z^2+h^2}, &\text{ otherwise} 
\end{cases}$ and following the same approach proposed in \cite{Galkin19}, the exact expression of the coverage probability can be obtained as 
\begin{align}
P_c=&\E_{Z_B}\big[\P({\rm SINR}>\tau \, | \, Z_B=z)\big]\nonumber\\
=&\sum\limits_{B=L,N,T} \E_{Z_B}\big[a_B(D_B)\,P_{c,B}(D_B)\,|\, Z_B=z\big]    \nonumber\\ =&\sum\limits_{B=L,N,T} \int\limits_{\R^+} a_B(d_B^B(z))\,P_{c,B}(d_B^B(z))\,f_{Z_B}(z)\,{\rm d}z \nonumber\\
=& \int\limits_0^\infty a_T({z}) \, P_{c,T}({z}) \, f_{Z_T}({z})\,{\rm d}{z} \nonumber\\
&+ \sum\limits_{M=L,N}\int\limits_{\max(0,r_e-r_u)}^\infty a_M\left(\sqrt{{z}^2+h^2}\right) \nonumber\\
&\times P_{c,M}\left(\sqrt{{z}^2+h^2}\right) \, f_{Z_M}({z})\,{\rm d}{z},
\end{align}
in which $D_B=d_B^B(Z_B)$ and the exact expressions of the conditional coverage probabilities are given by\footnote{In the particular case of Rayleigh fading channel ($m_B=1$), we can compute the conditional coverage probability as just
$P_{c,B}(D_B)=\exp\bigg({-\frac{\tau\,D_B^{\alpha_B}\,\sigma_n^2}{\xi_B}}\bigg) \,\mathcal{L}_{I,B}\bigg(\frac{\tau\,D_B^{\alpha_B}}{\xi_B},Z_B\bigg) $.}
\begin{align}
    P_{c,B}(z)=&\P\bigg(\frac{\xi_B\,H_B\,D_B^{-\alpha_B}}{J}>\tau\bigg)\nonumber\\
    =&\E_{D_B}\bigg[\P\bigg(H_B\,>\frac{\tau}{\xi_B}\,D_B^{\alpha_B}\,J\bigg)\bigg],
\end{align}
with $J=\sigma_n^2+I$.
By definition, the CCDF of the Gamma distribution is $\bar{F}_G(g)=\frac{\Gamma_u(m,m\,g)}{\Gamma(m)}$, where $\Gamma_u(m,m\,g)=\int\limits_{m\,g}^\infty t^{m-1}\,e^{-t}\,{\rm d}t$ is the upper incomplete Gamma function.
Let $\mu_B(D_B)=m_B\,\frac{\tau}{\xi_B}\,D_B^{\alpha_B}$, taking the expectation with respect to $J$ yields \cite{Alzenad19}
\begin{align}
    P_{c,B}=&\E_{D_B}\bigg[\E_J\left[\frac{\Gamma_u(m_B,\mu_B(D_B)\,J)}{\Gamma(m_B)}\right]\bigg]\nonumber\\
    \overset{(a)}{=}&\E_{D_B}\bigg[\E_J\bigg[e^{-\mu_B(D_B)\,J}\,\sum\limits_{k=0}^{m_B-1}\frac{(\mu_B(D_B)\,J)^k}{k!}\bigg]\bigg] \nonumber\\
    \overset{(b)}{=}&\E_{D_B}\bigg[\sum\limits_{k=0}^{m_B-1}\frac{(\mu_B(D_B))^k}{k!}\,\E_J\left[e^{-\mu_B(D_B)\,J}\,J^k\right]\bigg],
\end{align}
where (a) is from the definition $\frac{\Gamma_u(m,g)}{\Gamma(m)}=e^{-g}\sum\limits_{k=0}^{m-1}\frac{g^k}{k!}$, and (b) is obtained from the linearity of the expectation operator.
Taking into account that 
\begin{align}
\E_J\big[e^{-s\,J}\,J^k\big]=(-1)^k\,\frac{\partial^k}{\partial s^k}\,\mathcal{L}_J(s,z), \nonumber
\end{align}
where 
\begin{align}\mathcal L_{J}(s,z)=&\E\big[e^{-s\,J}\big]=\E\big[e^{-s\,I}\,e^{-s\sigma_n^2}\big]\nonumber\\
=&e^{-s\,\sigma_n^2}\,\E\big[e^{-s\,I}\big]=e^{-s\,\sigma_n^2}\,\mathcal L_{I}(s,z)\,,\nonumber
\end{align} the final expression is obtained. 
This, however, implies the need of computing higher order derivatives of the Laplace transform of the interference, with the number of terms to be computed directly depending on the value of $m_B$.

\section{Proof of Theorem \ref{thm:Pc_tilde}} \label{appx:Pc_tilde}
A tight bound can be applied to the CDF of the Gamma distribution in order to ease the computation of the conditional coverage probabilities provided in Theorem \ref{thm:Pc}. 
Let $\Gamma_l(m,m\,g)=\int\limits_0^{m\,g}t^{m-1}\,e^{-t}\,{\rm d}t$ denote the lower incomplete Gamma function, then the CDF of the Gamma distribution $F_{\Gamma}(g)=\frac{\Gamma_l(m,m\,g)}{\Gamma(m)}=1-\frac{\Gamma_u(m,m\,g)}{\Gamma(m)}$, can be bounded as
$$(1-e^{-\varepsilon_{1}\,m\,g})^{m}\leq\frac{\Gamma_l(m,m\,g)}{\Gamma(m)}\leq(1-e^{-\varepsilon_{2}\,m\,g})^{m},$$
where $\varepsilon_{1}=\begin{cases}
 1, & \text{if } m\geq1 \nonumber\\ (m!)^{-\frac{1}{m}}, & \text{if } m<1   \end{cases}$ and $\varepsilon_{2}=\begin{cases}
 (m!)^{-\frac{1}{m}}, & \text{if } m>1 \nonumber\\ 1, & \text{if } m\leq1   \end{cases}$. \\
Note that for $m=1$ the upper and the lower bounds become equal and thus $\frac{\Gamma_l(1,g)}{\Gamma(1)}=1-e^{-g}$. \\
It has been shown in \cite{Bai13} that the upper bound actually is a good approximation, hence we consider $\varepsilon_{2}=(m!)^{-\frac{1}{m}}$.\\
Let $\nu_B=m_B\,\frac{\tau}{\xi_B}\,D_B^{\alpha_B}$, the approximate conditional coverage probabilities can be derived as \cite{Alzenad19}
\begin{align}
    \tilde{P}_{c,B}=&\E_{J}\bigg[\frac{\Gamma_u(m_B,\nu_B\,J)^k}{\Gamma(m_B)}\bigg]
    =\E_{J}\bigg[1-\frac{\Gamma_l(m_B,\nu_B\,J)}{\Gamma(m_B)}\bigg]\nonumber\\
    \overset{(a)}{\approx}&1-\E_{J}\Big[\left(1-e^{-\varepsilon_{2,B}\,\nu_B\,J}\right)^{m_B}\Big] \nonumber\\
    \overset{(b)}{=}& 1-\E_{J}\bigg[\sum\limits_{k=0}^{m_B}\binom{m_B}{k}\,(1)^{m_B-k}\,(-e^{-\varepsilon_{2,B}\,\nu_B\,J})^k\bigg] \nonumber \\
    =&\E_{J}\bigg[\sum\limits_{k=1}^{m_B}\binom{m_B}{k}\,(-1)^{k+1}\,\exp(-k\,\varepsilon_{2,B}\,\nu_B\,J)\bigg] \nonumber\\
    \overset{(c)}{=}&\sum\limits_{k=1}^{m_B}\binom{m_B}{k}\,(-1)^{k+1}\,\E_{J}\big[\exp(-k\,\varepsilon_{2,B}\,\nu_B\,J)\big], 
\end{align}
where $(a)$ follows from the upper bound previously introduced, $(b)$ from the binomial theorem under the assumption that $m_B\in\mathbb{N}$, and $(c)$ from the linearity of the expectation operator. 
The final result in (\ref{eq:approx_PcB}) can be obtained by applying the definition of the conditional Laplace transform of the interference.

\bibliographystyle{IEEEtran}
\bibliography{Text.bib}

\vspace{1cm}

\begin{IEEEbiography}
[{\includegraphics[width=1in,height=1.25in,clip,keepaspectratio]{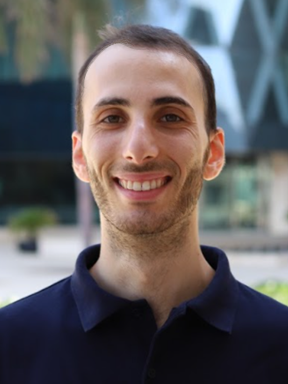}}]
{Maurilio Matracia}
is a Ph.D. student at King Abdullah University of Science and Technology (KAUST). 
He received his B.Sc. and M.Sc. degrees in Energy and Electrical Engineering from the University of Palermo (UNIPA), Italy, in 2017 and 2019, respectively.
His main research interest is stochastic geometry, with special focus on rural and emergency communications.
\end{IEEEbiography}


\begin{IEEEbiography}
[{\includegraphics[width=1in,height=1.25in,clip,keepaspectratio]{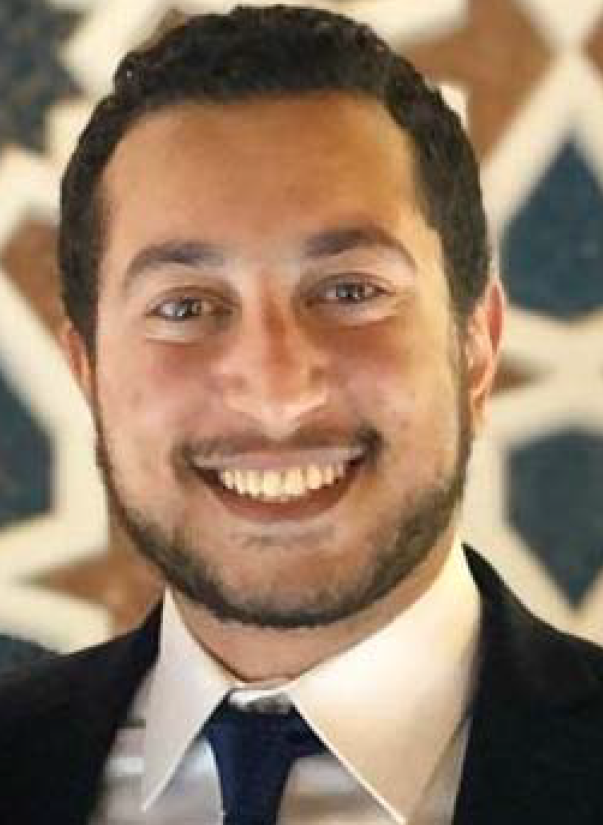}}]
{Mustafa A. Kishk}
[S'16, M'18] is a postdoctoral research fellow in the
communication theory lab at King Abdullah University of Science and Technology (KAUST). 
He received his B.Sc. and M.Sc. degree from Cairo University in 2013 and 2015, respectively, and his Ph.D. degree from Virginia Tech in 2018.
His current research interests include stochastic geometry, energy harvesting wireless networks, UAV-enabled communication systems, and satellite communications.
\end{IEEEbiography}


\begin{IEEEbiography}
[{\includegraphics[width=1in,height=1.25in,clip,keepaspectratio]{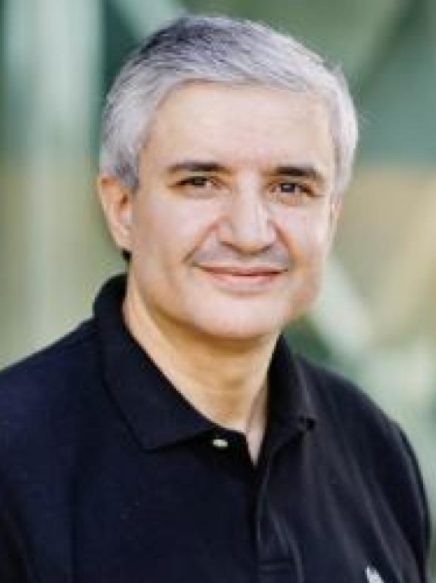}}]
{Mohamed-Slim Alouini} 
[S'94, M'98, SM'03, F'09] was born in Tunis, Tunisia.
He received the Ph.D. degree in Electrical Engineering from the California
Institute of Technology (Caltech), Pasadena, CA, USA, in 1998. 
He served as a faculty member in the University of Minnesota, Minneapolis, MN, USA, then in the Texas A$\&$M University at Qatar, Education City, Doha,
Qatar before joining King Abdullah University of Science and Technology
(KAUST), Thuwal, Makkah Province, Saudi Arabia as a Professor of Electrical Engineering in 2009. 
His current research interests include the modeling, design, and performance analysis of wireless communication systems.
\end{IEEEbiography}

\end{document}